\documentclass{book}
\usepackage{kenmax,xspace,epsfig}
\title{{\large\sc Model selection for inverse problems:} \\ 
       {\large\sc Best choice of basis function and model 
		 order selection}
\thanks{Presented at the 19th Int. worskhop on Bayesian 
and Maximum Entropy methods (MaxEnt 1999), Aug. 2-6, 1999, 
Boise, Idaho, USA}
}
\author{
A. Mohammad-Djafari \\[12pt]
Laboratoire des Signaux et Syst\`emes (CNRS-ESE-UPS),\\
Sup\'elec, Plateau de Moulon, 91192 Gif-sur-Yvette, France.\\[12pt]
E-mail: djafari@lss.supelec.fr
}
\def\babs{\begin{abstract}}             \def\eabs{\end{abstract}}
\def\barr{\begin{array}}                \def\earr{\end{array}}
\def\bcc{\begin{center}}                \def\ecc{\end{center}}

\def\bdes{\begin{description}}          \def\edes{\end{description}}
\def\bdoc{\begin{document}}             \def\edoc{\end{document}}
\def\ben{\begin{enumerate}}             \def\een{\end{enumerate}}
\def\beqn{\begin{eqnarray}}             \def\eeqn{\end{eqnarray}}
\def\beqnl#1{\beqn\label{#1}}           \def\eeqnl#1{\label{#1}\eeqn}
\def\beqnx{\begin{eqnarray*}}           \def\eeqnx{\end{eqnarray*}}
\def\bseqn{\begin{subeqnarray}}         \def\eseqn{\end{subeqnarray}}
\def\beq#1\eeq{\begin{equation}#1\end{equation}}
\def\bal#1\eal{\begin{align}#1\end{align}}
\def\balx#1\ealx{\begin{align*}#1\end{align*}}
\def\beqx{$$}                           \def\eeqx{$$}
\def\bfig{\protect\begin{figure}}       \def\efig{\protect\end{figure}}
\def\bfigx{\protect\begin{figure*}}     \def\efigx{\protect\end{figure*}}
\def\bfl{\begin{flushleft}}             \def\efl{\end{flushleft}}
\def\bfr{\begin{flushright}}            \def\efr{\end{flushright}}
\def\bit{\begin{itemize}}               \def\eit{\end{itemize}}
\def\bpic{\begin{picture}}              \def\epic{\end{picture}}
\def\bqu{\begin{quote}}                 \def\equ{\end{quote}}
\def\bqun{\begin{quotation}}            \def\equn{\end{quotation}}
\def\bsl{\begin{slide}}                 \def\esl{\end{slide}}
\def\btabb{\begin{tabbing}}             \def\etabb{\end{tabbing}}
\def\btabl{\begin{table}}               \def\etabl{\end{table}}
\def\btablx{\begin{table*}}             \def\etablx{\end{table*}}
\def\btab{\begin{tabular}}              \def\etab{\end{tabular}}
\def\btabu{\begin{tabular}}             \def\etabu{\end{tabular}}
\def\btabx{\begin{tabular*}}            \def\etabx{\end{tabular*}}
\def\bbib{\begin{thebibliography}{999}} \def\ebib{\end{thebibliography}}
\def\bver{\begin{verbatim}}             \def\ever{\end{verbatim}}

\def\bm#1{\mbox{\boldmath $#1$}}
\def\zerob{{\bm 0}}
\def\oneb{{\bm 1}}

\def\ab{{\bm a}}
\def\bb{{\bm b}}
\def\cb{{\bm c}}
\def\db{{\bm d}}
\def\eb{{\bm e}}
\def\fb{{\bm f}}
\def\gb{{\bm g}}
\def\hb{{\bm h}}
\def\ib{{\bm i}}
\def\jb{{\bm j}}
\def\kb{{\bm k}}
\def\lb{{\bm l}}
\def\mb{{\bm m}}
\def\nb{{\bm n}}
\def\ob{{\bm o}}
\def\pb{{\bm p}}
\def\qb{{\bm q}}
\def\rb{{\bm r}}
\def\sb{{\bm s}}
\def\tb{{\bm t}}
\def\ub{{\bm u}}
\def\vb{{\bm v}}
\def\wb{{\bm w}}
\def\xb{{\bm x}}
\def\yb{{\bm y}}
\def\zb{{\bm z}}

\def\Ab{{\bm A}}
\def\Bb{{\bm B}}
\def\Cb{{\bm C}}
\def\Db{{\bm D}}
\def\Eb{{\bm E}}
\def\Fb{{\bm F}}
\def\Gb{{\bm G}}
\def\Hb{{\bm H}}
\def\Ib{{\bm I}}
\def\Jb{{\bm J}}
\def\Kb{{\bm K}}
\def\Lb{{\bm L}}
\def\Mb{{\bm M}}
\def\Nb{{\bm N}}
\def\Ob{{\bm O}}
\def\Pb{{\bm P}}
\def\Qb{{\bm Q}}
\def\Rb{{\bm R}}
\def\Sb{{\bm S}}
\def\Tb{{\bm T}}
\def\Ub{{\bm U}}
\def\Vb{{\bm V}}
\def\Wb{{\bm W}}
\def\Xb{{\bm X}}
\def\Yb{{\bm Y}}
\def\Zb{{\bm Z}}

\def\alphab{\bm{\alpha}}
\def\betab{\bm{\beta}}
\def\deltab{\bm{\delta}}
\def\epsilonb{\bm{\epsilon}}
\def\gammab{\bm{\gamma}}
\def\omegab{\bm{\omega}}
\def\thetab{\bm{\theta}}
\def\xib{\bm{\xi}}
\def\lambdab{\bm{\lambda}}
\def\taub{\bm{\tau}}
\def\phib{\bm{\phi}}
\def\mub{\bm{\mu}}
\def\psib{\bm{\psi}}
\def\chib{\bm{\chi}}
\def\sigmab{\bm{\sigma}}

\def\Deltab{\bm{\Delta}}
\def\Lambdab{\bm{\Lambda}}
\def\Phib{\bm{\Phi}}
\def\Psib{\bm{\Psi}}
\def\Sigmab{\bm{\Sigma}}

\def\Ac{{\cal A}}
\def\Bc{{\cal B}}
\def\Cc{{\cal C}}
\def\Dc{{\cal D}}
\def\Ec{{\cal E}}
\def\Fc{{\cal F}}
\def\Gc{{\cal G}}
\def\Hc{{\cal H}}
\def\Ic{{\cal I}}
\def\Jc{{\cal J}}
\def\Kc{{\cal K}}
\def\Lc{{\cal L}}
\def\Mc{{\cal M}}
\def\Nc{{\cal N}}
\def\Oc{{\cal O}}
\def\Pc{{\cal P}}
\def\Qc{{\cal Q}}
\def\Rc{{\cal R}}
\def\Sc{{\cal S}}
\def\Tc{{\cal T}}
\def\Uc{{\cal U}}
\def\Vc{{\cal V}}
\def\Wc{{\cal W}}
\def\Xc{{\cal X}}
\def\Yc{{\cal Y}}
\def\Zc{{\cal Z}}

\def\wt#1{\widetilde{#1}}
\def\wh#1{\widehat{#1}}
\def\xh{\widehat{x}}
\def\thetah{\widehat{\theta}}
\def\betah{\widehat{\beta}}

\def\xbh{\widehat{\xb}}
\def\thetabh{\widehat{\thetab}}
\def\betabh{\widehat{\betab}}

\def\xbhk{\widehat{\xb}^{k}}
\def\thetahk{\widehat{\theta}^{k}}
\def\betahk{\widehat{\beta}^{k}}

\def\xbhkp{\widehat{\xb}^{k+1}}
\def\thetahkp{\widehat{\theta}^{k+1}}
\def\betahkp{\widehat{\beta}^{k+1}}

\def\thetabhk{\widehat{\thetab}^{k}}
\def\betabhk{\widehat{\betab}^{k}}

\def\thetabhkp{\widehat{\thetab}^{k+1}}
\def\betabhkp{\widehat{\betab}^{k+1}}

\def\thetamin{\theta_{\mbox{\tiny min}}}
\def\thetamax{\theta_{\mbox{\tiny max}}}
\def\betamin{\beta_{\mbox{\tiny min}}}
\def\betamax{\beta_{\mbox{\tiny max}}}

\def\ra{\rightarrow}
\def\la{\leftarrow}
\def\da{\downarrow}
\def\ua{\uparrow}

\def\Ra{\Rightarrow}
\def\La{\Leftarrow}
\def\Da{\Downarrow}
\def\Ua{\Uparrow}

\def\lra{\longrightarrow}
\def\lla{\longleftarrow}
\def\Lra{\Longrightarrow}
\def\Lla{\longleftarrow}

\def\lrarr{\leftrightarrow}
\def\Lrarr{\Leftrightarrow}
\def\udarr{\updownarrow}
\def\Uparr{\Updownarrow}

\def\d#1{\,\mbox{d}#1}
\def\dxdy{\d{x}\d{y}}
\def\dwxdwy{\d{\omega_x}\d{\omega_y}}
\def\dxdydz{\d{x}\d{y}\d{z}}

\def\disp#1{{\displaystyle #1}}
\def\diag#1{\mbox{diag}\left\{#1\right\}}

\def\Prob#1{\mbox{Pr}\left\{#1\right\}}
\def\var#1{\mbox{Var}\left\{#1\right\}}
\def\cov#1{\mbox{Cov}\left\{#1\right\}}
\def\corr#1{\mbox{Corr}\left\{#1\right\}}
\def\trace#1{\mbox{Tr}\left\{#1\right\}}
\def\rang#1{\mbox{rang}\left\{#1\right\}}
\def\det#1{\mbox{d\'et}\left\{#1\right\}}

\def\cosf{\cos \phi}
\def\sinf{\sin \phi}
\def\cost{\cos \theta}
\def\sint{\sin \theta}

\def\sgn{\mbox{sgn}}
\def\sinc{\mbox{sinc}}
\def\rect{\mbox{rect}}
\def\sincf#1{\mbox{sinc}\left(#1\right)}
\def\rectf#1{\mbox{rect}\left(#1\right)}
\def\trif#1{\mbox{tri}\left(#1\right)}
\def\xvec#1#2#3{\left\{#1_#2,\ldots,#1_#3\right\}}

\def\vx{\left[x_1,\ldots, x_n\right]^t}
\def\vz{\left[z_1,\ldots, z_n\right]^t}
\def\vw{\left[\omega_1,\ldots, \omega_n\right]^t}
\def\vxi{\left[\xi_1,\ldots, \xi_n\right]^t}
\def\iii{\int_{-\infty}^{+\infty}}
\def\izi{\int_{0}^{\infty}}
\def\izpi{\int_{0}^{\pi}}
\def\izdpi{\int_{0}^{2\pi}}
\def\intd{\int\kern-.8em\int}
\def\intt{\int\kern-.8em\int\kern-.8em\int}
\def\intg{\int\kern-1.1em\int}
\def\sumd{\mathop{\sum\sum}}

\def\sumi{\sum_{i=1}^{M}}
\def\sumj{\sum_{i=1}^{N}}
\def\sumk{\sum_{k=1}^{K}}
\def\sumn{\sum_{n=1}^{N}}
\def\summ{\sum_{m=1}^{M}}

\def\TA#1{{\cal A}\left\{ {#1} \right\}}
\def\TH#1{{\cal H}\left\{ {#1} \right\}}
\def\TP#1{{\cal P}\left\{ {#1} \right\}}
\def\TR#1{{\cal R}\left\{ {#1} \right\}}
\def\TRa#1{{\cal R}^{\dag}\left\{ {#1} \right\}}
\def\BR#1{{\cal B}\left\{ {#1} \right\}}
\def\TF#1{{\cal F}\left\{ {#1} \right\}}
\def\TFI#1{{\cal F}^{-1}\left\{ {#1} \right\}}
\def\TFn#1#2{{\cal F}_{#1}\left\{ {#2} \right\}}
\def\TFnI#1#2{{\cal F}_{#1}^{-1}\left\{ {#2} \right\}}
\def\Im#1{{\cal I}\mbox{m}\left(#1\right)}
\def\Ker#1{{\cal K}\mbox{er}\left(#1\right)}
\def\Imag#1{\mbox{Im}\left(#1\right)}
\def\Re#1{\mbox{Re}\left(#1\right)}
\def\expf#1{\exp\left[ {#1} \right]}

\def\dfdx#1#2{{\mbox{d} {#1}\over{\mbox{d} {#2}}}}
\def\dfdxd#1#2{{\mbox{d}^2 {#1}\over{\mbox{d} {#2}^2}}}
\def\dfdxt#1#2{{\mbox{d}^3 {#1}\over{\mbox{d} {#2}^3}}}
\def\dfdxn#1#2{{\mbox{d}^n {#1}\over{\mbox{d} {#2}^n}}}
\def\dfdxk#1#2{{\mbox{d}^k {#1}\over{\mbox{d} {#2}^k}}}

\def\dpdx#1#2{{{\partial {#1}\over \partial {#2}}}}
\def\dpdxd#1#2{{{\partial^2 {#1}}\over{\partial {#2}^2}}}
\def\dpdxdy#1#2#3{{{\partial ^2 {#1}}\over{\partial {#2} \partial {#3}}}}

\def\arg{\mbox{arg}}
\def\argmins#1#2{\mbox{arg}\min_{#1}\left\{{#2}\right\}}
\def\argmaxs#1#2{\mbox{arg}\max_{#1}\left\{{#2}\right\}}
\def\argmin#1#2{\mathop{\mbox{arg}\min}_{#1}\left\{{#2}\right\}}
\def\argmax#1#2{\mathop{\mbox{arg}\max}_{#1}\left\{{#2}\right\}}

\def\esp#1{\mbox{E}\left\{ #1 \right\}}
\def\espx#1#2{\mbox{E}_{#1}\left\{ #2 \right\}}

\def\wth#1{\widehat{\widetilde{\phantom{#1}}}\!\!\!\! #1}

\def\lrf{L_{r,\phi}}
\def\fw{\widehat{f}(\omegab)}
\def\fthwxi{\wth{f}(\Omega,\xib)}
\def\fthwfi{\wth{f}(\Omega,\phi)}
\def\ftrfi{\widetilde{f}(r,\phi)}

\def\fwxwy{\widehat{f}(\omega_x, \omega_y)}
\def\wxpwy{(\omega_x \, x + \omega_y \, y)}

\def\wtx{\omegab^t \cdot \xb}
\def\ejwtx{\exp\left[j \omegab^t \cdot \xb\right]}
\def\xitx{\xib^t \cdot \xb}

\def\ftrxi{\widetilde{f}(r,\xib)}
\def\ent{-\int p(x) \, \ln p(x) \d{x}}

\def\mean#1{\left< #1 \right>}
\def\slnhn{\sum_{n=1}^N \lambda_n h_n(\rb)}
\def\slngn{\sum_{n=1}^N \lambda_n g_n(\rb)}
\def\smngn{\sum_{n=1}^N \mu_n g_n(\rb)}
\def\slmhm{\sum_{m=1}^N \lambda_m h_m(\rb)}
\def\vlambda{\bm{\lambda} = [\lambda_1,\ldots,\lambda_n]}

\def\apriori{{\em a priori} }
\def\aposteriori{{\em a posteriori} }

\def\titre#1{\bcc{\Large\bf #1}\ecc}

\def\AMD{Ali Mohammad--Djafari}
\def\LSSa{Laboratoire des Signaux et Syst\`emes 
(CNRS--ESE--UPS) \\ 
\'Ecole Sup\'erieure d'\'Electricit\'e \\ 
Plateau de Moulon, 91192 Gif sur Yvette Cedex, France.}

\def\ME{maximum entropy}
\def\pdf{probability distribution function}
\def\lm{Lagrange multipliers}
\def\fix#1{\phi _#1(x)}
\def\fin{\fix n}
\def\fik{\fix k}
\def\fiz{\fix 0}
\def\sfinz{\sum_{n=0}^N \lambda_n \, \fin}
\def\sfinu{\sum_{n=1}^N \lambda_n \, \fin}
\def\bl{\bm{\lambda}}
\def\bd{\bm{\delta}}
\def\blz{\bl ^0}
\def\gnl{G _n(\bl)}
\def\gnlz{G _n(\blz)}
\def\un{n=1,\dots, N}
\def\nn{n=0,\dots, N}

\def\finn{\fin , \nn}
\def\esfinz{\exp\,\left[ -\sfinz \right] }
\def\esfinu{\exp\,\left[ -\sfinu \right] }
\def\esxm{\exp\,\left[ -\sum_{m=0}^N \lambda_m \, x^m \right] }
\def\efin{\esp \fin }
\def\zl{Z(\bl)}
\def\finxi{\phi _n(x_i)}
\def\snfinxi{\sum_{n=1}^N \lambda_n \finxi}
\def\esnfinxi{\exp \left[ - \snfinxi \right]}
\def\smfinxi{\sum_{i=1}^M \finxi}

\def\ejnw{\exp \left( -j n \omega_0 x \right) }
\def\eejnw{\mbox{E} \left\lbrace \ejnw \right\rbrace}

\def\signed#1{{\unskip\nobreak\hfil\penalty50\hskip2em\mbox{}
\nobreak\hfil\tt#1\parfillskip=0pt \finalhyphendemerits=0 \par}}

\def\uncatcodespecials{\def\do##1{\catcode`##1=12 }\dospecials}
\def\listing#1{\par\begingroup\setupverbatim\input#1 \endgroup}
\newcount\lineno
\def\setupverbatim{\tt \lineno=0
 \obeylines \uncatcodespecials \obeyspaces
 \everypar{\advance\lineno by1 \llap{\sevenrm\the\lineno\ \ }}}
{\obeyspaces\global\let =\ }

\def\defined{\stackrel{\mbox{def}}{=}}
\def\str{\stackrel}

\def\ER{\mbox{I\kern-.25em R}}
\def\EC{\mbox{C\kern-.8em C}}
\def\EZ{\mbox{Z\kern-.55em Z}}
\def\EN{\mbox{N\kern-.8em N}}

\def\singles{
 \abovedisplayskip 12pt plus 3pt minus 9pt
 \belowdisplayskip 12pt plus 3pt minus 9pt
 \abovedisplayshortskip 0pt plus 3pt
 \belowdisplayshortskip 7pt plus 3pt minus 4pt
 \baselineskip 14.4pt
 \lineskip 1pt
 \lineskiplimit 0pt}
\def\oneandhalf{
 \abovedisplayskip 18pt plus 3pt minus 9pt
 \belowdisplayskip 18pt plus 3pt minus 9pt
 \abovedisplayshortskip 0pt plus 3pt
 \belowdisplayshortskip 9.333pt plus 3pt
 \baselineskip 20pt
 \lineskip 2pt
 \lineskiplimit 1pt}

\def\double{
 \abovedisplayskip 24pt plus 3pt minus 9pt
 \belowdisplayskip 24pt plus 3pt minus 9pt
 \abovedisplayshortskip 0pt plus 3pt
 \belowdisplayshortskip 12pt plus 3pt
 \baselineskip 27pt
 \lineskip 3pt
 \lineskiplimit 2pt}

\def\dadb{\d{\alpha}\d{\beta}}

\def\ffbox#1{\fbox{\mbox{\vbox{#1}}}}

\def\rot{\mbox{rot}}
\def\case#1#2#3#4{
    \left\{
           \begin{array}{ll}
            {\displaystyle #1} & {\displaystyle #2} \cr 
            {\displaystyle #3} & {\displaystyle #4}
           \end{array}
    \right. }

\def\beqnarr#1&#2&#3\\#4&#5&#6\eeqnarr{
    \left\{
           \begin{array}{lcl}
            {\displaystyle #1} & #2 & {\displaystyle #3} \\ 
            {\displaystyle #4} & #5 & {\displaystyle #6} 
           \end{array}
    \right. }

\def\pyx{p(\yb|\xb)}
\def\pxy{p(\xb|\yb)}

\def\ie{{\em i.e.}}
\def\unsdpi{\left(\frac{1}{2\pi}\right)}
\def\unspi{\left(\frac{1}{\pi}\right)}
\def\up{\uppercase}
\def\zjm{z_{j-1}}
\def\zjp{z_{j+1}}
\def\fxyp{f(x,y)=\left\{
\barr{ll} 1 & (x,y)\in P\\ 0 & (x,y)\not\in P\earr
\right.}

\def\FigDir{}

\def\bm#1{\mbox{\boldmath $#1$}}
\def\wt#1{\widetilde{#1}}
\def\wh#1{\widehat{#1}}
\def\ER{\mbox{I\kern-.25em R}}
\def\intg{\int\kern-1.1em\int}
\def\d#1{\,\mbox{d}#1}

\def\det#1{\left| #1 \right|}
\def\Tr#1{\mbox{Tr}\left\{ #1 \right\}}
\def\rem#1{}

\def\xb{\bm{x}}
\def\yb{\bm{y}}
\def\epsilonb{\bm{\epsilon}}
\def\Ab{\bm{A}}
\def\Ib{\bm{I}}
\def\phib{\bm{\phi}}
\def\psib{\bm{\psi}}
\def\thetab{\bm{\theta}}

\def\xbz{\bm{x}_0}

\def\phipsi{(\phi,\psi)}
\def\phibpsib{(\phib,\psib)}

\def\xbh{\widehat{\xb}}
\def\ybh{\widehat{\yb}}
\def\phibh{\widehat{\phib}}
\def\psibh{\widehat{\psib}}
\def\phibhpsibh{(\widehat{\phib}, \widehat{\psib})}
\def\kh{\widehat{k}}
\def\lh{\widehat{l}}
\def\ih{\widehat{i}}
\def\jh{\widehat{j}}
\def\phih{\widehat{\phi}}
\def\psih{\widehat{\psi}}
\def\lambdah{\widehat{\lambda}}
\def\phihpsih{(\widehat{\phi}, \widehat{\psi})}

\def\Pyy{\bm{P}_{y}}
\def\Pxx{\bm{P}_{x}}
\def\Pbb{\bm{P}_{b}}

\def\Pxay{\bm{P}_{x|y}}
\def\Pyay{\bm{P}_{y|y}}
\def\xmap{\wh{\bm{x}}_{\hbox{MAP}}}
\def\Pxh{\wh{\bm{P}}}

\def\model{\yb=\Ab\xb+\epsilonb}

\def\pyxphipsickl{p(\yb, \xb, \phib, \psib \,|\, k, l)}
\def\pycxphikl{p(\yb \,|\, \xb, \phib, k, l)}
\def\pxcpsikl{p(\xb \,|\, \psib, k, l)}
\def\pphickl{p(\phib \,|\, k, l)}
\def\ppsickl{p(\psib \,|\, k, l)}
\def\pk{p(k)}
\def\pl{p(l)}

\def\pyxcphipsikl{p(\yb,\xb \,|\, \phib, \psib, k, l)}
\def\pycphipsikl{p(\yb \,|\, \phib, \psib, k, l)}
\def\pxcyphipsikl{p(\xb \,|\, \yb, \phib, \psib, k, l)}
\def\pycphipsikl{p(\yb \,|\, \phib, \psib, k, l)}
\def\pycpsikl{p(\yb \,|\, \psib, k, l)}
\def\pycphikl{p(\yb \,|\, \phib, k, l)}
\def\pyphipsickl{p(\yb, \phib, \psib \,|\, k, l)}
\def\pypsickl{p(\yb, \psib \,|\, k, l)}
\def\pyphickl{p(\yb, \phib \,|\, k, l)}
\def\pphipsicykl{p(\phib, \psib \,|\, \yb, k, l)}
\def\pphipsickl{p(\phib, \psib \,|\, k, l)}
\def\pphicykl{p(\phib \,|\, \yb, k, l)}
\def\pphickl{p(\phib \,|\, k, l)}
\def\ppsicykl{p(\psib \,|\, \yb, k, l)}
\def\ppsickl{p(\psib \,|\, k, l)}
\def\pyckl{p(\yb \,|\, k, l )}
\def\pycl{p(\yb \,|\, l )}
\def\py{p(\yb)}

\def\pkcphipsiy{p(k \,|\, \yb, \phib, \psib)}
\def\pkcyl{p(k \,|\, \yb, l)}
\def\plcy{p(l \,|\, \yb)}

\def\pxcyphipsikh{p(\xb \,|\, \yb, \phibh, \psibh, \kh)}
\def\pphipsicykh{p(\phib, \psib \,|\, \yb, \kh)}
\def\pphicykh{p(\phib \,|\, \yb, \kh)}
\def\ppsicykh{p(\psib \,|\, \yb, \kh)}
\def\pthetacykh{p(\thetab \,|\, \yb, \kh)}
\def\pkcylh{p(k \,|\, \yb, \lh)}
\def\pycklh{p(\yb \,|\, k, \lh )}
\def\pyclh{p(\yb \,|\, \lh )}

\def\pxcyphipsiklh{p(\xb \,|\, \yb, \phibh, \psibh, \kh, \lh)}
\def\pphipsicyklh{p(\phib, \psib \,|\, \yb, \kh, \lh)}
\def\pphicyklh{p(\phib \,|\, \yb, \kh, \lh)}
\def\ppsicyklh{p(\psib \,|\, \yb, \kh, \lh)}
\def\pthetacyklh{p(\thetab \,|\, \yb, \kh, \lh)}

\def\pyxphipsickls{p(\yb, \xb, \phi, \psi \,|\, k, l)}
\def\pyxphipsikls{p(\yb, \xb, \phi, \psi, k, l)}
\def\pycxtkls{p(\yb \,|\, \xb, \phi, k, l)}
\def\pxcpsikls{p(\xb \,|\, \psi, k, l)}
\def\pphickls{p(\phi \,|\, k, l)}
\def\ppsickls{p(\psi \,|\, k, l)}

\def\pyphilambdackls{p(\yb,\phi,\lambda|k,l)}

\def\pycftkls{p(\yb \,|\, \fb, \phi, k, l)}
\def\pfcpsikls{p(\fb \,|\, \psi, k, l)}

\def\pyxcphipsikls{p(\yb,\xb \,|\, \phi, \psi, k, l)}
\def\pycphipsikls{p(\yb \,|\, \phi, \psi, k, l)}
\def\pxcyphipsikls{p(\xb \,|\, \yb, \phi, \psi, k, l)}
\def\pycphipsikls{p(\yb \,|\, \phi, \psi, k, l)}
\def\pyphipsickls{p(\yb, \phi, \psi \,|\, k, l)}
\def\pycpsikls{p(\yb \,|\, \psi, k, l)}
\def\pycphikls{p(\yb \,|\, \phi, k, l)}
\def\pypsickls{p(\yb, \psi \,|\, k, l)}
\def\pyphickls{p(\yb, \phi \,|\, k, l)}
\def\pphipsicykls{p(\phib, \psi \,|\, \yb, k, l)}
\def\pphipsickls{p(\phi, \psi \,|\, k, l)}
\def\pphicykls{p(\phi \,|\, \yb, k, l)}
\def\pphickls{p(\phi \,|\, k, l)}
\def\ppsicykls{p(\psi \,|\, \yb, k, l)}
\def\ppsickls{p(\psi \,|\, k, l)}
\def\pkcphipsiys{p(k \,|\, \yb, \phi, \psi)}
\def\pkcys{p(k \,|\, \yb)}

\def\pxcyphipsiklhs{p(\xb \,|\, \yb, \phih, \psih, \kh, \lh)}
\def\pphipsicyklhs{p(\phi, \psi \,|\, \yb, \kh, \lh)}
\def\pphicyklhs{p(\phi \,|\, \yb, \kh, \lh)}
\def\ppsicyklhs{p(\psi \,|\, \yb, \kh, \lh)}

\def\kmax{k_{max}}
\def\lmax{l_{max}}

\begin{document}
\maketitle
\begin{abstract}
A complete solution for an inverse problem needs five main steps: choice of  
basis functions for discretization, determination of the order of the model, 
estimation of the hyperparameters, estimation of the solution, and finally, 
caracterisation of the proposed solution. Many works have been done for 
the three last steps. The two first have been neglected for a while, 
in part due to the complexity of the problem. 
However, in many inverse problems, particularly when the number of data is 
very low, a good choice of the basis functions and a good selection of the 
order become primordial.  
In this paper, we first propose a complete solution whithin a Bayesian 
framework. Then, we apply the proposed method to an inverse elastic 
electron scattering problem. 
\end{abstract}

\section{Introduction}
In a very general linear inverse problem, the relation between the data 
$\yb=[y_1,\cdots,y_m]^t$ and the unknown function $f(.)$ is 
\beq \label{model}
y_i= \intg h_i(\rb) \, f(\rb) \d{\rb}, \quad i=1,\cdots,m, 
\eeq
where $h_i(\rb)$ is the system response for the data $y_i$. 
We assume here that $h_i(\rb)$ are known perfectly. 
The first step for any numerical processing is the choice of a 
basis function $b_j(\rb)$ and an order $k$, in such a way to 
be able to write 
\beq
f(\rb) = \sum_{j=1}^k x_j \, b_j(\rb).  
\eeq
This leads to 
\beq
\yb = \Ab \xb + \epsilonb
%y_i = \sum_{j=1}^k  A_{i,j} \, x_j, \quad i=1,\cdots,m  
\eeq 
with $\yb=[y_1,\cdots,y_m]^t$, $\xb=[x_1,\cdots,x_k]^t$ and 
\beq
A_{i,j} = \intg h_i(\rb) \, b_j(\rb) \d{\rb}, 
\quad i=1,\cdots,m, \, j=1,\cdots,k
\eeq
where $\epsilonb=[\epsilon_1,\cdots,\epsilon_m]^t$ 
represents the errors (both the measurement noise and 
the modeling and the approximation related to the numerical 
computation of matrix elements $A_{i,j}$). 
Even, when the choice of the basis functions $b_i(\rb)$ and the 
model order $k$ is fixed, obtaining a good estimate for $\xb$ needs 
other assumptions about the noise $\epsilonb$ and about $\xb$ itself. 
The Bayesian approach provides a coherent and complete 
framework to handle the random nature 
of $\epsilonb$ and the \apriori incomplete knowledge of 
$\xb$.  

The first step in a Bayesian approach is to assign the prior 
probability laws 
$p(\yb \,|\, \xb, \phib, k, l)
=p_{\epsilon}(\yb-\Ab\xb \,|\, \phib, k, l)$, 
$\pxcpsikl$, $\pphickl$ and $\ppsickl$, 
where $p_{\epsilon}(\yb-\Ab\xb|\phi, k, l)$ is 
the probability law of the noise, 
and $(\phib, \psib)$ the hyperparameters of the problem. 
Note that $\xb$ represents the unknown parameters, 
$k=\hbox{dim}(\xb)$ is the order of the model,  
$m=\hbox{dim}(\yb)$ is the number of the data 
and $l$ is an index to a particular choice of basis functions.   
Note that the elements of the 
matrix $\Ab$ depend on the choice of the basis functions. 
However, to simplify the notations, 
we do not write this dependance explicitely. 
We assume that 
we have to select one set $l$ of basis functions among a finite set 
(indexed by $[1:\lmax]$) of them.  
Thus, for a given $l\in[1,\lmax]$ and a given model order 
$k\in[1,\kmax]$, 
and using the mentionned prior laws, we define the joint 
probability law  
\beq
\pyxphipsickl = \pycxphikl \, \pxcpsikl \, \pphickl \, \ppsickl.
\eeq
From this probability law, we obtain,  
either by integration or by summation, any marginal law, 
and any \aposteriori probability law using the Bayes rule. 

What we propose in this paper is to consider the following problems: 
\bit
\item Parameter estimation:
\beq
\xbh = \argmax{\xb}{\pxcyphipsiklh}
\eeq 
where 
\beq
\pxcyphipsikl = \pyxcphipsikl \, / \, \pycphipsikl, 
\eeq
\beq 
\pyxcphipsikl = \pycxphikl \, \pxcpsikl 
\eeq
and
\beq
\pycphipsikl = \intg \pyxcphipsikl \d{\xb}.
\eeq

\item Hyperparameter estimation:
\beq
\phibhpsibh = \argmax{\phibpsib}{\pphipsicyklh}
\eeq 
where
\beq
\pphipsicykl = \pyphipsickl \, / \, \pyckl
\eeq
and
\beq
\pyckl = \intg \intg \pyphipsickl \d{\phib} \d{\psib}. 
\eeq

\item Model order selection:
\beq
\kh = \argmax{k}{\pkcylh} %= \argmax{k}{\pk \, \pycklh /\pyclh}
\eeq 
where
\beq
\pkcyl = \pyckl \, \pk / \, \pycl
\eeq
and 
\beq
\pycl = \sum_{k=1}^{\kmax} \pyckl \, \pk. 
\eeq

\item Basis function selection:
\beq
\lh = \argmax{l}{\plcy} %= \argmax{l}{\pycl \, \pl \, / \, \py}
\eeq 
where
\beq
\plcy = \pycl \, \pl / \, \py
\eeq
and
\beq
\py = \sum_{l=1}^{\lmax} \pycl \, \pl.  
\eeq

\item Joint parameter, hyperparameter, model order and basis 
function estimation:
\beq
(\xbh, \phibh, \psibh, \kh, \lh) 
= \argmax{(\xb,\phib,\psib,k,l)}{\pyxphipsickl \, \pk \, \pl}. 
\eeq 
\eit

As it can be easily seen, the first problem is, in general, 
a well posed problem and the solution can be computed, 
either analytically or numerically. 
The others (excepted the last) need integrations. 
These integrals can be done analytically 
only in the case of Gaussian laws. In other cases, one can either use 
a numerical integration (either deterministic or stochastic) 
or to resort to approximations such as the Laplace's method which allows 
to obtain a closed-form expression for the criterion to optimise. 

Here, we consider these problems 
for the particular case of Gaussian prior laws: 
\beqn 
\pycxtkls &=& {\cal N}\left(\Ab\xb, \frac{1}{\phi} \Ib\right) 
= \left(2\pi/\phi\right)^{-m/2} 
\expf{-\frac{1}{2} \phi \, \|\yb-\Ab\xb\|^2} \label{pyx} \\ 
\pxcpsikls  &=& {\cal N}\left(\bm{0}, \frac{1}{\psi} \Ib\right) 
= \left(2\pi/\psi\right)^{-k/2} 
\expf{-\frac{1}{2} \psi \, \| \xb \|^2} \label{px}
\eeqn
where $\frac{1}{\phi}$ and $\frac{1}{\psi}$ are respectively the 
variance of the noise and the parameters. 

\section{Parameter estimation}

First note that in this special case we have
\beq
\pyxcphipsikls = \left(2\pi/\phi\right)^{-m/2} 
\left(2\pi/\psi\right)^{-k/2} 
\expf{-\frac{1}{2} \phi \, \|\yb-\Ab\xb\|^2
-\frac{1}{2} \psi \, \| \xb\|^2}. 
\eeq
Integration with respect to $\xb$ can be done analytically 
and we have:
\beq
\pycphipsikls  = \intg \pyxcphipsikls \d{\xb} 
= {\cal N}\left(\bm{0}, \Pyy\right),  
\eeq
with
\beq \label{Pyy}
\Pyy=\frac{1}{\psi} \Ab\Ab^t +\frac{1}{\phi} \Ib
=\frac{1}{\psi} (\Ab\Ab^t+\lambda\Ib)
\hbox{~~and~~} \lambda=\frac{\psi}{\phi}.    
\eeq
It is then easy to see that the {\em a posteriori} law 
of $\xb$ is also Gaussian: 
\beq 
\pxcyphipsikls = {\cal N}\left(\xbh, \Pxh\right) 
\hbox{~with~~} \Pxh=\frac{1}{\phi} (\Ab^t\Ab+\lambda \Ib)^{-1} 
\hbox{~and~~} \xbh=\phi \Pxh \Ab^t \yb. 
\eeq
Thus the parameter estimation in this case is straightforward:  
\beq
\xbh = \argmax{\xb} \pxcyphipsikls = \argmin{\xb}{J_1(\xb)}, 
\eeq
with
\beq \label{J1}
J_1(\xb)= \|\yb-\Ab\xb\|^2+\lambda\| \xb \|^2, 
\eeq
which is a quadratic function of $\xb$. The solution is then 
a linear function of the data $\yb$ and is given by
\beq
\xbh= \Kb(\lambda) \, \yb \quad\hbox{with}\quad 
\Kb(\lambda)=(\Ab^t\Ab+\lambda \Ib)^{-1}\Ab^t.
\eeq 

\section{Hyperparameter estimation}

For the hyperparameter estimation problem we note that:
\beqn
\pphipsicykls  
&=&  \frac{\pphickls \, \ppsickls}{\pyckl} \,\, \pycphipsikls 
\nonumber \\ 
&=& \frac{\pphickls \, \ppsickls}{\pyckl} (2\pi)^{-m/2} \det{\Pyy}^{-1/2} 
\expf{-\frac{1}{2} \yb^t \Pyy^{-1}\yb}. 
\eeqn
Thus, the hyperparameter estimation problem becomes:
\beq
\phihpsih = \argmax{\phipsi}{\pphipsicyklhs} 
= \argmin{\phipsi}{J_2(\phi,\psi)}
\eeq 
where
\beq \label{J2a}
J_2(\phi,\psi)= -\ln \pphickls -\ln \ppsickls 
+\frac{1}{2} \, \ln \det{\Pyy} + \frac{1}{2} \, \yb^t \Pyy^{-1} \yb. 
\eeq 
Unfortunately, in general, there is not an analytical expression 
for the solution, but this optimization can be done numerically. 
Many works have been investigated to perform this optimization 
appropriately for 
particular choices of $\pphickls$ and $\ppsickls$. 
Among the others, we may note the choice of improper prior laws 
such as Jeffry's prior 
$\pphickls \propto \frac{1}{\phi}$ and $\ppsickl \propto \frac{1}{\psi}$ 
or proper prior laws of uniform 
$\pphickls =\frac{1}{\phi_{\max}-\phi_{\min}}$ and 
$\ppsickls =\frac{1}{\psi_{\max}-\psi_{\min}}$ or still 
the proper Gamma prior laws.

One main issue with improper prior laws is the existence of the solution, 
because $\pphipsicykls$ may even not have a maximum or its maximum 
can be located at the border of the domain of variation of 
$(\phi,\psi)$. 
Here, we propose to use the following proper Gamma priors : 
\beqn
p(\phi) &=& {\cal G}(\alpha_1,\beta_1)
\propto \phi^{(\alpha_1-1)}\expf{-\beta_1\phi} 
\lra \esp{\phi}=\alpha_1/\beta_1
\\ 
p(\psi) &=& {\cal G}(\alpha_2,\beta) 
\propto \psi^{(\alpha_2-1)}\expf{-\beta_2\psi}
\lra \esp{\psi}=\alpha_2/\beta_2. 
\eeqn
With these priors, we have
\beq \label{J2b}
J_2(\phi,\psi)
= (1-\alpha_1)\ln\phi+(1-\alpha_2)\ln\psi+\beta_1\phi+\beta_2\psi
+\frac{1}{2} \, \ln \det{\Pyy} + \frac{1}{2} \, \yb^t \Pyy^{-1} \yb. 
\eeq 

The second main issue is the numerical optimization. 
Many works have been done on this subject. Among the others we 
can mention those who try to integrate out one of the two 
parameters directly or after some transformation. 
For example transforming  $(\phi,\psi)\lra(\phi,\lambda)$ and 
using the identities 
\beq
\det{\Ab\Ab^t+\lambda\Ib}
=\lambda^{m-k}\det{\Ab^t\Ab+\lambda\Ib} 
\eeq
and
\beq
(\Ab\Ab^t+\lambda\Ib)^{-1}
=\frac{1}{\lambda}(\Ib-\Ab\Kb(\lambda))
\eeq
we have
\beq
\ln \det{\Pb_y}= -m \ln\phi -k\ln\lambda
+\ln\det{\Ab^t\Ab+\lambda\Ib} \\ 
\eeq
and
\beq
\yb^t\Pb_y^{-1}\yb
=\phi \, \yb^t(\Ib-\Ab\Kb(\lambda))\yb
=\phi \, \yb^t (\yb-\Ab\wh{\xb})
=\phi \, \yb^t (\yb-\wh{\yb}).
\eeq
Then, we obtain   
\beq \label{J2c}
\barr{ll}
J_2(\phi,\psi)=
& (1-\alpha_1-\frac{m-k}{2})\ln\phi
 +(1-\alpha_2-\frac{k}{2})\ln\psi
 + \beta_1\phi+\beta_2\psi \\ 
&+ \frac{1}{2} \, \ln \det{\Ab^t\Ab+\frac{\psi}{\phi}\Ib} 
 + \frac{\phi}{2} \, \yb^t (\yb-\wh{\yb}). 
\earr
\eeq 
or
\beq \label{J2d}
\barr{ll}
J_2(\phi,\lambda)=
& (2-\alpha_1-\alpha_2-\frac{m}{2})\ln\phi
 +(1-\alpha_2-\frac{k}{2})\ln\lambda
 + \beta_1\phi+\beta_2\phi\lambda \\ 
&+ \frac{1}{2} \, \ln \det{\Ab^t\Ab+\lambda\Ib} 
 + \frac{\phi}{2} \, \yb^t (\yb-\wh{\yb}). 
\earr
\eeq 
For fixed $\lambda$, equating to zero the derivative of this 
expression with respect to $\phi$ has an explicite solution 
which is 
\beq \label{Phic}
\dpdx{J_2(\phi,\lambda)}{\phi}=0 \lra 
\phi=(\frac{m}{2}+\alpha_1+\alpha_2-2)
\,/\, 
\left[
\beta_1+\lambda\beta_2+\frac{1}{2}\yb^t(\yb-\wh{\yb}) 
\right] 
\eeq 
Putting this expression into $J_2$ we obtain a criterion depending 
only on $\lambda$ which can be optimized numerically. 
In addition, it is possible to integrate out $\phi$ to obtain 
$p(\lambda | \yb, k,l)$, but the expression is too complex  
to write. 

%The third issue concerns the integration with respect to $\xb$ and, 
%more particularly, the integration with respect to $\phi$ and $\psi$. 

\rem{
We can show that subject to some constraints on the parameters 
$\alpha_1$, $\alpha_2$, $\beta_1$ and $\beta_2$ we can be insured 
to have a maximum. 
Indead, as we will see, in some cases, it is possible to obtain 
simple fix-point or gradient type algorithms for the optimization 
problems. 
Finally, with choice, some of the integrals have explicite solutions. 
{\tt A COMPLETER}
\beq
\left\{
\barr{l}
\dpdx{J_2}{\phi} = \\ 
\dpdx{J_2}{\phi} = 
\earr
\right.
\eeq
}

\section{Joint estimation}
One may try to estimate all the unknowns simultaneously by
\beq
(\xbh, \phih, \psih, \kh, \lh) 
= \argmax{(\xb,\phi,\psi,k,l)}{p(\xb,\phi,\psi, k, l | \yb}
= \argmin{(\xb,\phi,\psi,k,l)}{J_3(\xb,\phi,\psi,k,l)} 
\eeq 
where 
\beq \label{J3}
\barr{ll}
J_3(\xb,\phi,\psi,k,l)= 
& -\ln \pk -\ln \pl 
  -(\frac{m}{2}+\alpha_1-1) \ln \phi \\ 
& -(\frac{k}{2}+\alpha_2-1) \ln \psi 
  + \phi \left(\beta_1 + \frac{1}{2} \|\yb-\Ab\xb\|^2 \right)
  + \psi \left(\beta_2 + \frac{1}{2} \|\xb\|^2 \right)  
\earr
\eeq
The main advantage of this criterion is that we obtain explicit solutions 
for $\xb$, $\phi$ and $\psi$ by equating to zero the derivatives of $J_3(\xb,\phi,\psi,k,l)$ with respect to them: 
\beq
\left\{\barr{l} 
\xbh= (\Ab^t\Ab+\lambda \Ib)^{-1} \Ab^t \yb, \qquad\hbox{with}\qquad 
\lambda=\phi / \psi ; \\ 
\phih=(\frac{m}{2}+\alpha_1-1) / 
\left(\beta_1 + \frac{1}{2} \|\yb-\Ab\xbh\|^2 \right) ; \\ 
\psih=(\frac{k}{2}+\alpha_2-1) / 
\left(\beta_2 + \frac{1}{2} \|\xbh\|^2 \right). \\ 
\earr\right.
\eeq
We can not obtain closed form expressions for $\kh$ and $\lh$ which 
depend on the particular choice for $\pk$ and $\pl$.  
These relations suggest an iterative algorithm such as: 

\medskip\noindent
\begin{minipage}{12cm}
\fbox{\hspace*{1em}\hbox{\vbox{
\noindent{\tt 
\centerline{Joint MAP estimation algorithm 1}\\ 
for $l=1:\lmax$ \\ 
\hspace*{1em} for $k=1:\kmax$ \\ 
\hspace*{2em} compute the elements of the matrix $\Ab$; \\ 
\hspace*{2em} initialize $\lambdah=\lambda_0$; \\ 
\hspace*{2em} repeat until convergency: 
\[
\barr{l} 
~~\xbh= (\Ab^t\Ab+\lambdah \Ib)^{-1} \Ab^t \yb;  \\ 
\left\{\barr{l}
\phih=(\frac{m}{2}+\alpha_1-1) 
/ \left(\beta_1 + \frac{1}{2} \|\yb-\Ab\xbh\|^2 \right); \\ 
\psih=(\frac{k}{2}+\alpha_2-1) 
/ \left(\beta_2 + \frac{1}{2} \|\xbh\|^2 \right) \\ 
\earr\right. 
\lra \lambdah=\phih / \psih
\earr
\]
\hspace*{2em} end \\ 
\hspace*{2em} compute $J(k,l)=J_3(\xbh,\phih,\psih,k,l)$; \\ 
\hspace*{1em} end \\ 
end \\ 
choose the best model and the best order by \\  
\centerline{$(\lh,\kh)=\argmin{(k,l)}{J(k,l)}$} \\  
}
}}}
\end{minipage}

Note however that, for fixed $\xb$, $\phi$ and $\psi$, the criteria 
$J_3$ in (\ref{J3}) or $J_5$ in (\ref{J5}) are mainly linear functions 
of $k$ if we choose a uniform law for $\pk$. 
This means that we may not have a minimum for these criteria as a 
function of $k$. The choice of the prior $\pk$ is then important. 
One possible is the following:
\beq
p(k)=\left\{\barr{ll}
\frac{2(\kmax-k)}{\kmax(\kmax-1)} & 1\le k < \kmax \\ 
0                                 & k > \kmax
\earr\right.
\eeq
which is a decreasing function of $k$ in the range $k\in[1,\kmax]$ and 
zero elsewhere. This choice may insure the existence of a minimum if 
$\kmax$ is choosed appropriately. For $\pl$ we propose to choose 
a uniform law, because we do not want to give any favor to any model. 

Another algorithm can be obtained if we replace the expression 
of $\xbh$ into 
$J_3$ to obtain a criterion depending only on $(\phi,\psi)$
\beq \label{J4}
\barr{ll}
J_4(\phi,\psi,k,l)=
&-\ln\pk-\ln\pl
-(\frac{m}{2}+\alpha_1-1) \ln \phi 
-(\frac{k}{2}+\alpha_2-1) \ln \psi 
\\ 
& + \phi \left(\beta_1 + \frac{1}{2}
  \|\yb-\wh{\yb}(\lambda)\|^2 \right)
  + \psi \left(\beta_2 + \frac{1}{2} 
  \|\wh{\xb}(\lambda)\|^2 \right)
\earr
\eeq
or on $(\phi,\lambda)$:
\beq \label{J5}
\barr{ll}
J_5(\phi,\lambda,k,l)=
& -\ln\pk-\ln\pl 
  -(\frac{m+k}{2}+\alpha_1+\alpha_2-2) \ln \phi 
  -(\frac{k}{2}+\alpha_2-1) \ln \lambda  \\ 
& + \phi \left(\beta_1 + \frac{1}{2}
  \|\yb-\wh{\yb}(\lambda)\|^2 \right)
  + (\lambda\phi) \left(\beta_2 + \frac{1}{2} 
  \|\wh{\xb}(\lambda)\|^2 \right)
\earr
\eeq
and then optimize it with respect to them. 
In the second case, we can again obtain first $\phi$ and put it's 
expression 
\beq \label{Phi7}
\wh{\phi}=
\left(\frac{m+k}{2}+\alpha_1+\alpha_2-2\right) \,/\,
\left[
(\beta_1 + \frac{1}{2}\|\yb-\wh{\yb}(\lambda)\|^2)
+ \lambda(\beta_2 + \frac{1}{2} \|\wh{\xb}(\lambda)\|^2)
\right]
\eeq
in the criterion to obtain another criterion depending only on 
$\lambda$ and optimize it numerically. 
This gives the following algorithm:

\noindent 
\begin{minipage}{12cm}
\fbox{\hspace*{1em}\hbox{\vbox{
\noindent{\tt 
\centerline{Joint MAP estimation algorithm 2}\\ 
for $l=1:\lmax$ \\ 
\hspace*{1em} for $k=1:\kmax$ \\ 
\hspace*{2em} compute the elements of the matrix $\Ab$; \\ 
\hspace*{2em} for $\lambda\in 10^{[-8:1:4]}$  \\ 
\hspace*{3em}  compute 
                 $\xbh= (\Ab^t\Ab+\lambda \Ib)^{-1}\Ab^t\yb$ 
					  and $\ybh=\Ab\xbh$ \\ 
\hspace*{3em}  compute $\phih$ using (eq.~\ref{Phi7}) \\ 
\hspace*{3em}  compute $J(\lambda)=J_5(\phih,\lambda,k,l)$ (eq.~\ref{J5}) \\ 
\hspace*{2em} end \\ 
\hspace*{2em} choose $\lambdah=\argmin{\lambda}{J(\lambda)}$ \\ 
\hspace*{2em} compute $\xbh= (\Ab^t\Ab+\lambdah \Ib)^{-1} \Ab^t \yb$; \\ 
\hspace*{2em} compute $\phih$ using (eq.~\ref{Phi7}); \\ 
\hspace*{2em} compute $J(k,l)=J_5(\phih,\lambdah,k,l)$ (eq.~\ref{J5}) \\ 
\hspace*{1em} end \\ 
end \\ 
choose the best model and the best order by \\  
 \centerline{$(\lh,\kh)=\argmin{(k,l)}{J(k,l)}$}  \\ 
}
}}}
\end{minipage}

\section{Model order selection}
The model order selection  
\beq
\kh = \argmax{k} \pkcyl = \argmin{k}{J_6(k)}
\eeq 
with
\beq \label{J6}
J_6(k)=-\ln \pk - \ln \pyckl
\eeq
needs one more integration 
\beq 
\pyckl   = \intd \pyphipsickls \d{\phi} \d{\psi}.  
\eeq
or
\beq 
\pyckl   = \intd \pyphilambdackls \d{\phi} \d{\lambda}.  
\eeq
where $\pyphilambdackls\propto\expf{-J_2(\phi,\lambda)}$ given by (\ref{J2d}).
As we mentionned in preceeding section, these integrations can only be down 
numerically. A good approximation can be obtained using the following:
\beq \label{pyckl}
\pyckl   = \int \int \pyphipsickls \d{\phi} \d{\psi} 
\simeq \sum_i \sum_j p(\yb|\phi_j,\psi_i,k,l) 
\eeq
where $\{\phi_j\}$ and $\{\psi_i\}$ are samples generated using the prior 
laws $p(\phi)$ and $p(\psi)$. 

\section{Best basis or model selection}
The model selection  
\beq
\lh = \argmax{l} \plcy = \argmin{l}{J_7(l)}
\eeq 
with
\beq \label{J7}
J_7(l)=-\ln \pl - \ln \pycl
\eeq 
does not need any more integration, but only one summation. 
Choosing $\pl$ uniform and making the same previous approximations we have
\beq \label{J7b}
J_7(l)=-\ln \sum_{k=1}^{\kmax} \pyckl \, \pk.
\eeq

\section{Proposed algorithms}
Based on the equations (\ref{J7}), (\ref{pyckl}), (\ref{J6}), (\ref{J2c}) 
and (\ref{J2d}) we propose the following algorithm:
\\ 
\begin{minipage}{12cm}
\fbox{\hspace*{1em}\hbox{\vbox{
\noindent{\tt 
\centerline{Marginal MAP estimation algorithm 2}\\ 
Generate a set of samples $\{\phi_j\}$ drawn from $p(\phi)$  \\ 
Generate a set of samples $\{\psi_i\}$ drawn from $p(\psi)$  \\ 
for $l=1:\lmax$ \\ 
\hspace*{1em} for $k=1:\kmax$ \\ 
\hspace*{2em} compute the elements of the matrix $\Ab$; \\ 
\hspace*{2em} for $\phi\in\{\phi_j\}$  \\ 
\hspace*{3em} for $\psi\in\{\psi_i\}$  \\ 
\hspace*{4em} compute 
                 $\lambda=\phi/\psi$, \quad 
					  $\xbh= (\Ab^t\Ab+\lambda \Ib)^{-1}\Ab^t\yb$ 
					  and $\ybh=\Ab\xbh$ \\ 
\hspace*{4em} compute $p_{\psi}(i,j,k,l)=\expf{-J_2(\phi_j,\psi_i)}$ 
                (eq.~\ref{J2c}) \\ 
\hspace*{3em} end \\ 
\hspace*{3em} normalize 
  $p_{\psi}(i,j,k,l)=p_{\psi}(i,j,k,l) \,/\, \sum_i p_{\psi}(i,j,k,l)$ \\ 
\hspace*{3em} compute 
  $p_{\phi}(j,k,l)=\sum_i p_{\psi}(i,j,k,l)$  \\ 
\hspace*{2em} end \\ 
\hspace*{2em} normalize 
  $p_{\phi}(j,k,l)=p_{\phi}(j,k,l) \,/\, \sum_j p_{\phi}(j,k,l)$ \\ 
\hspace*{2em} compute $p_k(k,l)=\sum_j p_{\phi}(j,k,l)$  \\ 
\hspace*{1em} end \\ 
\hspace*{1em} normalize $p_k(k,l)=p_k(k,l) \,/\, \sum_k p_k(k,l)$  \\ 
\hspace*{1em} compute $p_l(l)=\sum_k p_k(k,l)$  \\ 
end \\ 
normalize $p_l(l)=p_l(l) \,/\, \sum_l p(l)$ \\ 
choose the best model by $\lh=\argmax{l}{p_l(l)}$ \\ 
choose the best model order by $\kh=\argmax{k}{p_k(k,\lh)}$ \\ 
choose the best value for $\phi=\phi_{\jh}$ with 
 $\jh=\argmax{j}{p_{\phi}(j,\lh,\kh)}$ \\ 
choose the best value for $\psi=\psi_{\ih}$ with 
 $\ih=\argmax{i}{p_{\psi}(i,\jh,\lh,\kh)}$ \\ 
compute $\lambdah=\phih/\psih$ \\ 
compute the elements of the matrix $\Ab$ for $l=\lh$ and $k=\kh$ \\ 
compute $\xbh= (\Ab^t\Ab+\lambdah \Ib)^{-1}\Ab^t\yb$. \\ 
}
 }}}
\end{minipage}

\rem{ 
Up to know, we focused on the estimation $\xb$ for a given model $l$ 
and a given model order $k$. But, the final quantity of interest is 
not $\xb$ but $f(\rb)$ or at least its values for some discrete 
set of values of $\{\rb_n, n=1,\ldots,N\}$. 
In a one dimensional case, if we note by 
$\fb=\{f(r_n), r_n=(n-1)\Delta r, \, n=1,\ldots,N\}$, then we have 
\beq
\fb=\Bb\xb 
\eeq
where $\Bb$ is a $(N\times k)$ matrix with elements $\Bb_{nj}=b_j(r_n)$. 
We have then to express all the preceeding equations as a function of 
$\fb$ in place of $\xb$. 
This can be done very easily by replacing 
everywhere $p(\yb|\xb,...)$ by $p(\yb|\Bb\fb,...)$ and 
$p_x(\xb|...)$ by $p_f(\fb|...)=\det{\Bb\Bb^t}^{-1} p_x(\Bb\fb|...)$. 
For example the equations (\ref{pyx}) and (\ref{px}) have to be 
replaced by 
\beq 
\pycftkls={\cal N}\left(\Ab\Bb\fb, \frac{1}{\phi} \Ib\right) 
= \left(2\pi/\phi\right)^{-m/2} 
\expf{-\frac{1}{2} \phi \, \|\yb-\Ab\Bb\fb\|^2} \label{pyf}
\eeq
and
\beq
\pfcpsikls={\cal N}\left(\bm{0}, \frac{1}{\psi} \Bb\Bb^t\right) 
= \left(2\pi/\psi\right)^{-k/2} \det{\Bb\Bb^t}^{-1/2} 
\expf{-\frac{1}{2} \psi \, \| \Bb\xb \|^2} \label{pf}
\eeq
}

%\newpage
\section{Application: Electron scattering data inversion}
Elastic electron scattering provides a mean of determining the
charge density of a nucleus, $\rho(r)$, from the experimentally
determined charge form factor, $F(q)$.
The connection between the charge density and
the cross section is well understood and in plane wave Born
approximation $F(q)$ is just the Fourier transform of $\rho(r)$
which for the case of even-even nuclei, which we shall consider,
is simply given by
\begin{equation}\label{Fc}
F(q) = 4\pi\int_0^{\infty }r^2\,J_0(q r)\rho(r)\d{r}
\end{equation}
where $J_0$ is the spherical Bessel function of zero order and $q$
is the absolute value of the three momentum transfer.  
\rem{Given that
the experimental measurements are performed over a limited range at
a finite number of values of the momentum transfer $q$, a unique
determination of $\rho(r)$ is not possible since the resulting
inverse problem is ill posed.
}

We applied the proposed method with the following usual 
discretization procedure:
\begin{equation}\label{Exp0}
\rho(r) =\left\{
\begin{array}{ll}
\sum_{j=1}^{k} x_j\, b_j(r) & r\leq R_{c}\\
0                           & r > R_{c}
\end{array}
         \right.
\end{equation}
which results in
\begin{equation}\label{Fcexp0}
F(q)= 4\pi\sum_{j=1}^{k} x_j \int_0^{R_{c}} r^2\,J_0(q r) \, b_j(r)\d{r}
\end{equation}
and
\begin{equation}\label{Discrete0}
\bm{y}=\bm{A}\bm{x} +\bm{\epsilon}
\end{equation}
where $\bm{x}$ is a vector containing the coefficients
$\{x_{j}, j=1,\cdots,k\}$, $\bm{y}$ is a vector containing the
form factor data $\{F(q_{i}), i=1,\cdots,m\}$
and $\Ab$ an $(m\times k)$ matrix containing the coefficients
$A_{i,j}$ given by 
\begin{equation}\label{Amn}
A_{i,j}=4\pi\int_0^{R_{c}} r^2\,J_0(q_{i} r)\, b_j(r)\d{r}.
\end{equation}

To compute $A_{i,j}$ we define a discretization step $\Delta r=R_c/N$, 
a vector $\rb=\{r_n=(n-1)\Delta r, n=1,\cdots,N\}$, 
a $(N\times k)$ matrix $\Bb$ with elements $B_{n,j}=b_j(r_n)$, 
a $(m\times N)$ matrix $\Cb$ with elements 
$C_{i,n}=(4\pi \Delta r) r_n^2 J_0(q_i r_n)$ such that we have $\Ab=\Cb\Bb$. 
Note also that when the vector $\bm{x}$ is determined, we can compute 
$\bm{\rho}=\{\rho(r_n), \, n=1,\cdots,N\}$ by $\bm{\rho}=\bm{B} \bm{x}$. 

%\subsection{Numerical experiments to test and compare different methods}
To test the proposed methods, we used the following simulation procedure:
\bit
\item Select a model type $l$ and an order $k$ and generate the 
the matrixes $\Bb$, $\Cb$ and $\Ab$, and for a random set of parameters 
$\xb$ generate the data $\yb=\Ab\xb$. 

\item Add some noise $\epsilonb$ on $\yb$ to obtain $\yb=\Ab\, \xb+\epsilonb$;

\item Compute the estimates $\lh$, $\kh$, $\xbh$, $\ybh=\Ab\, \xbh$ and 
$\wh{\bm{\rho}}=\Bb\xbh$ and compare them with $l$, $k$, $\xb$, 
$\yb=\Ab\xb$ and $\bm{\rho}=\Bb\xb$. 
\eit
We choosed the following basis functions: 
\bit 
\item $l=1$~: \quad $b_j(r)= J_0(q_j r)$--- 
This is a natural choice due to the integral kernel and the orthogonality 
property of the bessel functions. 

\item $l=2$~: \quad $b_j(r)= \hbox{sinc}(q_j r)$---  
This choice is also natural due to the orthogonality and the limited support 
hypothesis for the function $\rho(r)$. 

\item $l=3$~: \quad $b_j(r)= \expf{- \frac{1}{2}(q_j r)^2}$--- 
This choice can account for the positivity of the function $\rho(r)$ 
if $\{x_j\}$ are constrained to be positive. 

\item $l=4$~: \quad $b_j(r)= \expf{- \frac{1}{2} (q_j r)^2} J_0(q_j r)$--- 
This choice combines the first and the third properties. 

\item $l=5$~: \quad $b_j(r)= 1/(\hbox{cosh}(q_j r))$--- 
This choice has the same properties of the third one.  

\item $l=6$~: \quad $b_j(r)= 1/(1+(q_j r)^2)$--- 
This choice has the same properties of the third one.  
\eit

In all these experiments we choosed $k=6$, $m=20$, $N=100$, $R_c=8$ and 
$q_i = i \pi/R_c$. 
\rem{This choice is interesting, 
because, thanks to the orthogonality property of the bessel functions, 
the matrix $\Ab$ is diagonal and we have an explicite solution for $\xb$ 
given by:
\begin{equation}\label{Coeff}
x_j=\frac{F(q_j)}{2\pi R_{c}^{3} \left[J_{1}(q_j R)\right]^{2}}.
\end{equation}
}
The following figures show typical solutions. Figures 1 and 2 show the details 
of the procedure for the case $l=1$. Figures 3, 4 and 5 show the results for 
the cases $l=1$ to $l=6$. 

\newpage
%\begin{figure}
\begin{tabular}[b]{@{}l@{}l@{}} 
\begin{tabular}[b]{@{}l@{}} 
\epsfxsize=90mm\epsfysize=60mm\epsfbox{\FigDir 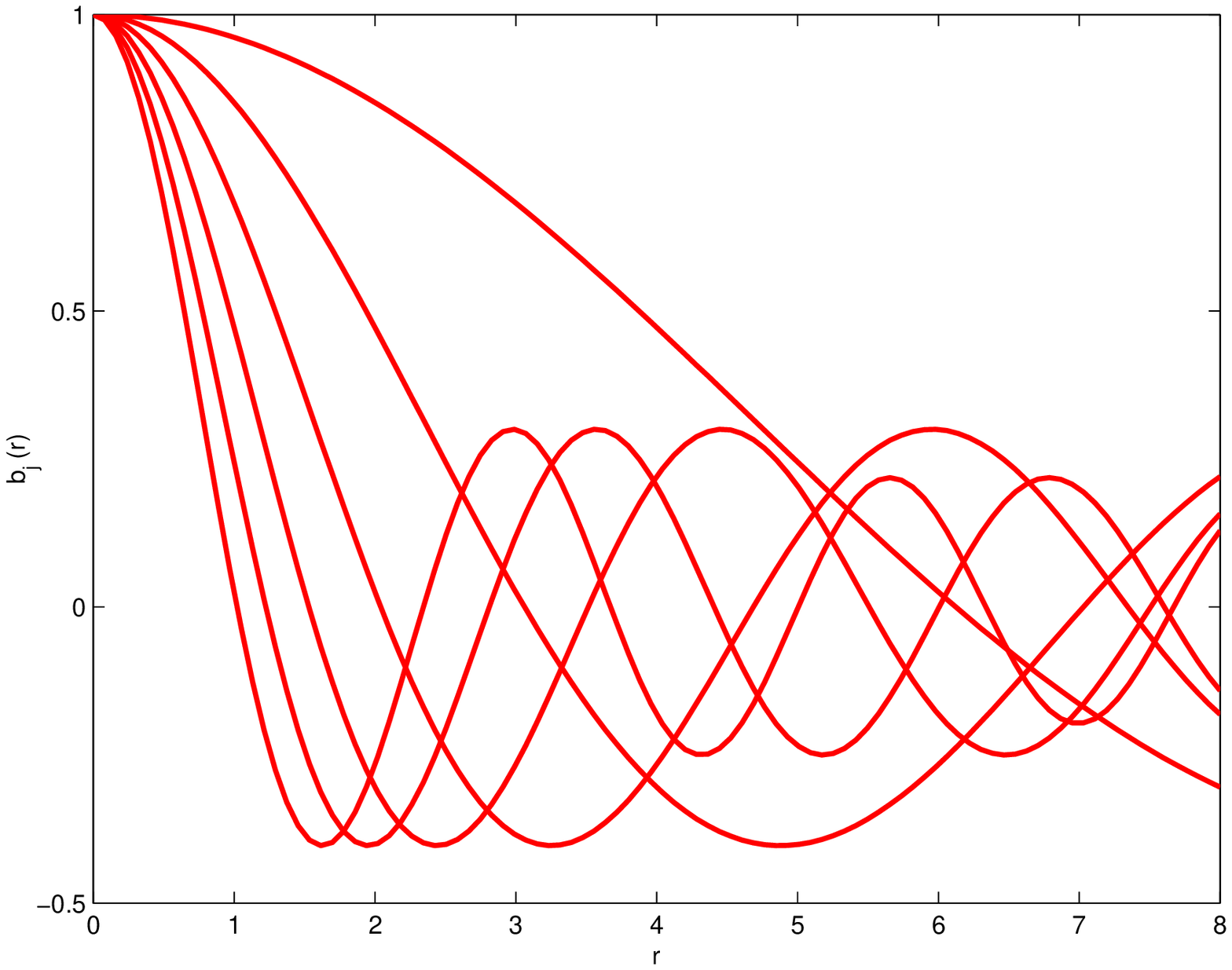} 
\end{tabular} 
&
\begin{tabular}[b]{@{}l@{}} 
$\disp{~b_j(r)}$\\ $~j=1,\ldots,k=6$  \\ ~ \\ ~ \\ ~\\ ~\\ ~\\ ~\\
\end{tabular}
\\
\begin{tabular}[b]{@{}l@{}} 
\epsfxsize=90mm\epsfysize=60mm\epsfbox{\FigDir 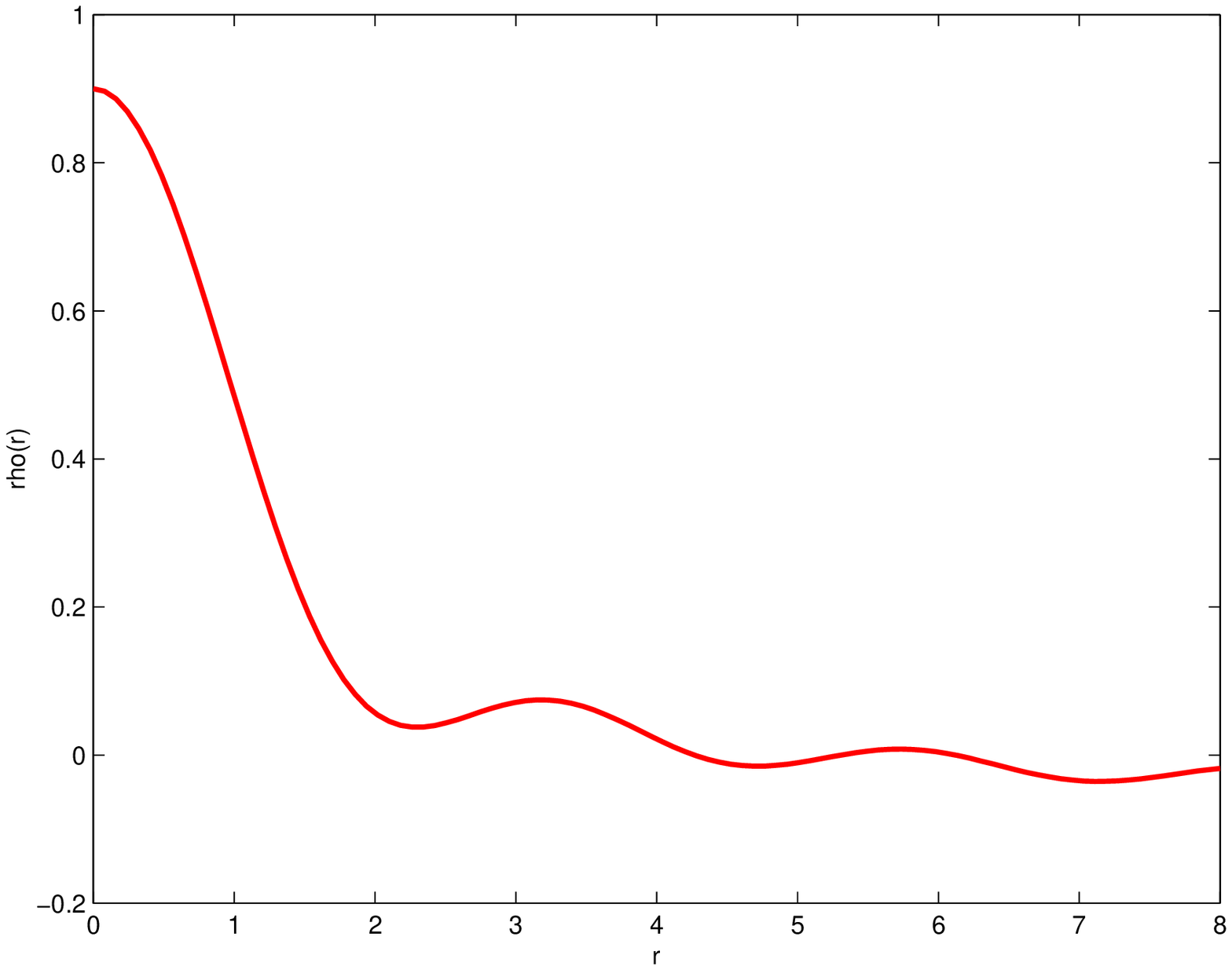} 
\end{tabular} 
&
\begin{tabular}[b]{@{}l@{}} 
$\disp{\rho(r)=\sum_{j=1}^{k} x_j \, b_j(r)}$  \\ 
with \\ 
$x_1=x_2=\cdots=x_6=1$ \\ ~ \\ ~\\ ~\\ ~\\ ~\\ ~\\
\end{tabular}
\\
\begin{tabular}[b]{@{}l@{}} 
\epsfxsize=90mm\epsfysize=60mm\epsfbox{\FigDir 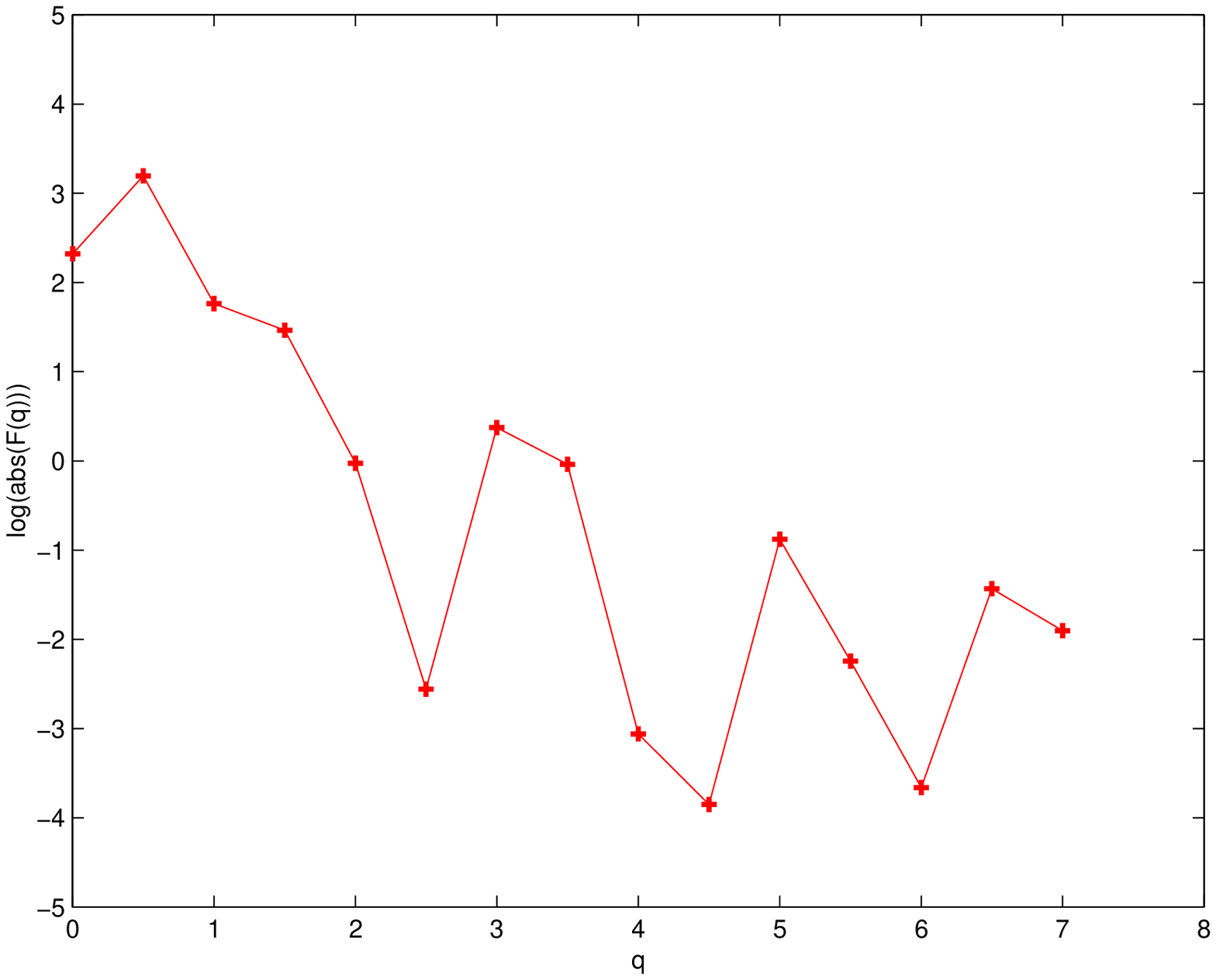} 
\end{tabular} 
&
\begin{tabular}[b]{@{}l@{}} 
$\disp{F(q_i)=\sum_{j=1}^{k} A_{ij} \, x_j}$\\ 
$\disp{A_{ij}=\int r^2 J_0(q_i r) \, b_j(r) \d{r}}$ \\
$~i=1,\ldots,m=14$ \\ 
$~j=1,\ldots,k=6$ \\   \\ ~ \\ ~ \\ ~\\
\end{tabular}
\end{tabular}
\\
\centerline{Fig. 1:\, a) basis functions $b_j(r)$,\quad 
b) $\rho(r)$,\quad 
c) data $F(q_i)$ in a logarithmic scale.}
%\caption{a) basis functions,\quad b) $\rho(r)$,\quad c) data $F(q_i)$.}
%\end{figure}

\newpage
%\begin{figure}
\begin{tabular}[b]{@{}l@{}l@{}} 
\begin{tabular}[b]{@{}l@{}} 
\epsfxsize=90mm\epsfysize=60mm\epsfbox{\FigDir 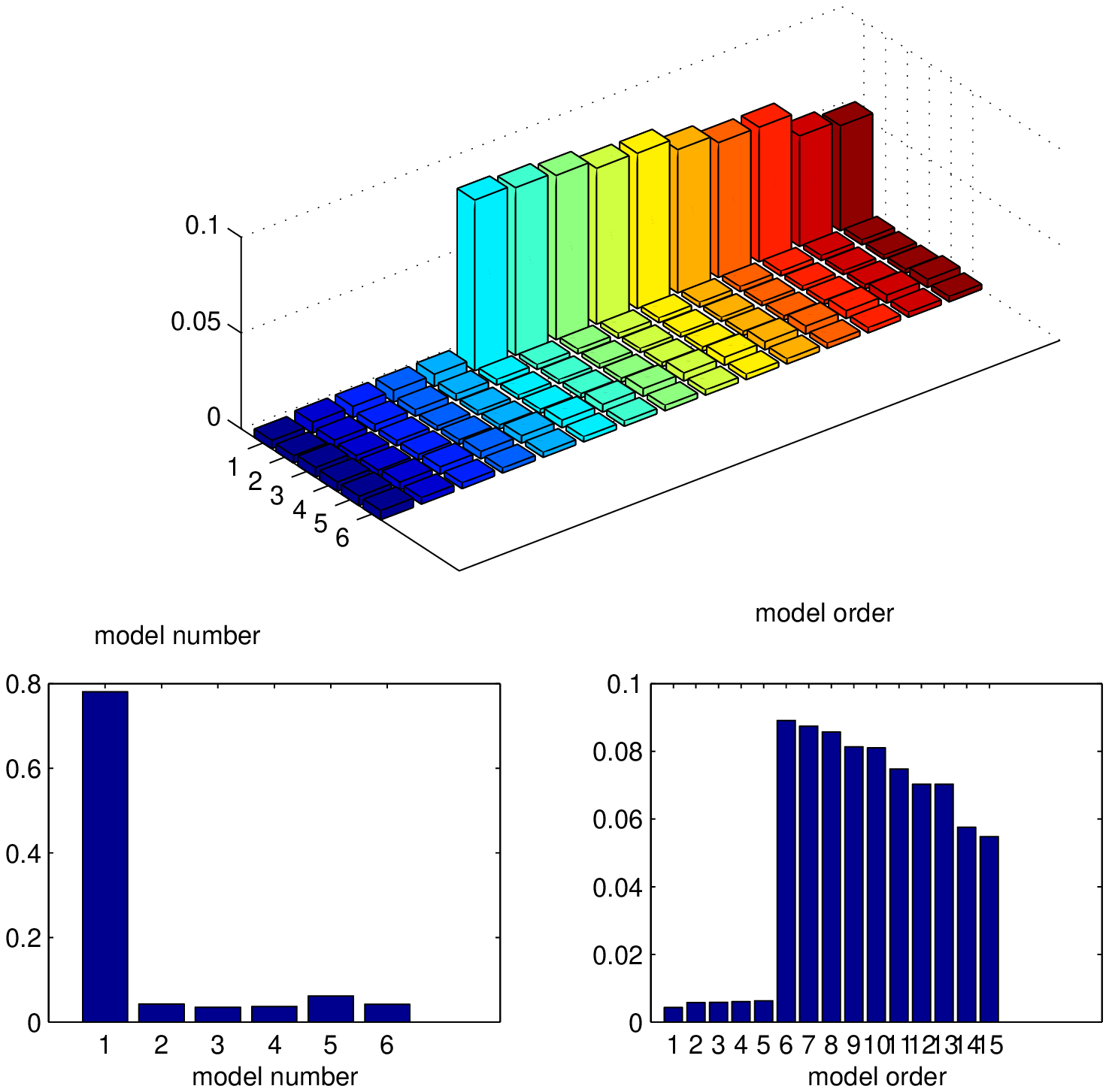} 
\end{tabular} 
&
\begin{tabular}[b]{@{}l@{}} 
$p(k,l)$ \\ ~ \\ ~ \\ ~\\ ~\\ ~\\ 
$p(l)$ and $p(k|\lh)$ \\ ~\\  ~\\ ~\\
\end{tabular}
\\
\begin{tabular}[b]{@{}l@{}} 
\epsfxsize=90mm\epsfysize=60mm\epsfbox{\FigDir 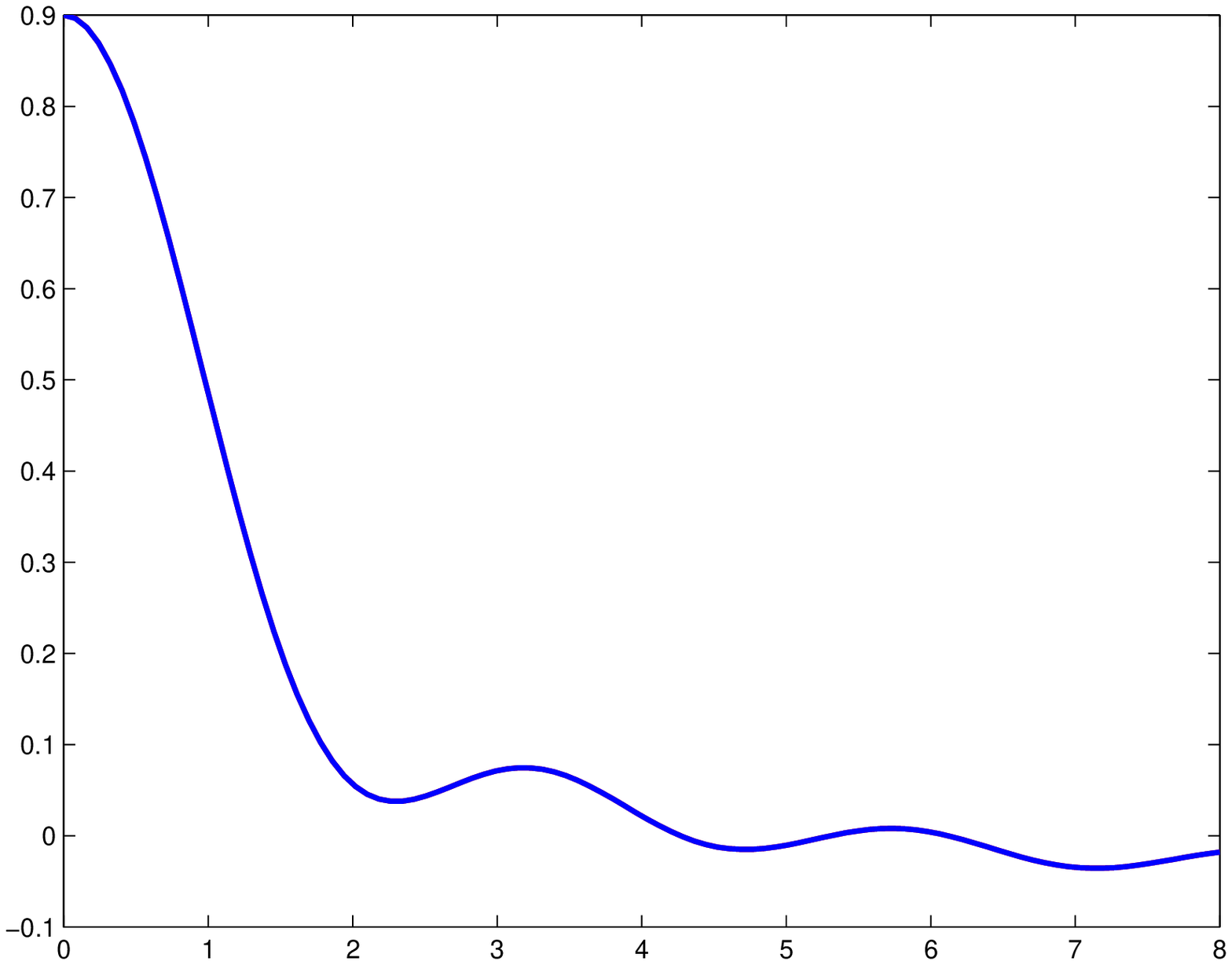} 
\end{tabular} 
&
\begin{tabular}[b]{@{}l@{}} 
$\disp{\wh{\rho}(r)=\sum_{j=1}^k \wh{x}_j \, b_j(r)}$ \\ 
and \\ 
$\disp{\rho(r)=\sum_{j=1}^k x_j \, b_j(r)}$ 
\\ ~ \\ ~ \\ ~\\ ~\\ 
\end{tabular}
\\
\begin{tabular}[b]{@{}l@{}} 
\epsfxsize=90mm\epsfysize=60mm\epsfbox{\FigDir 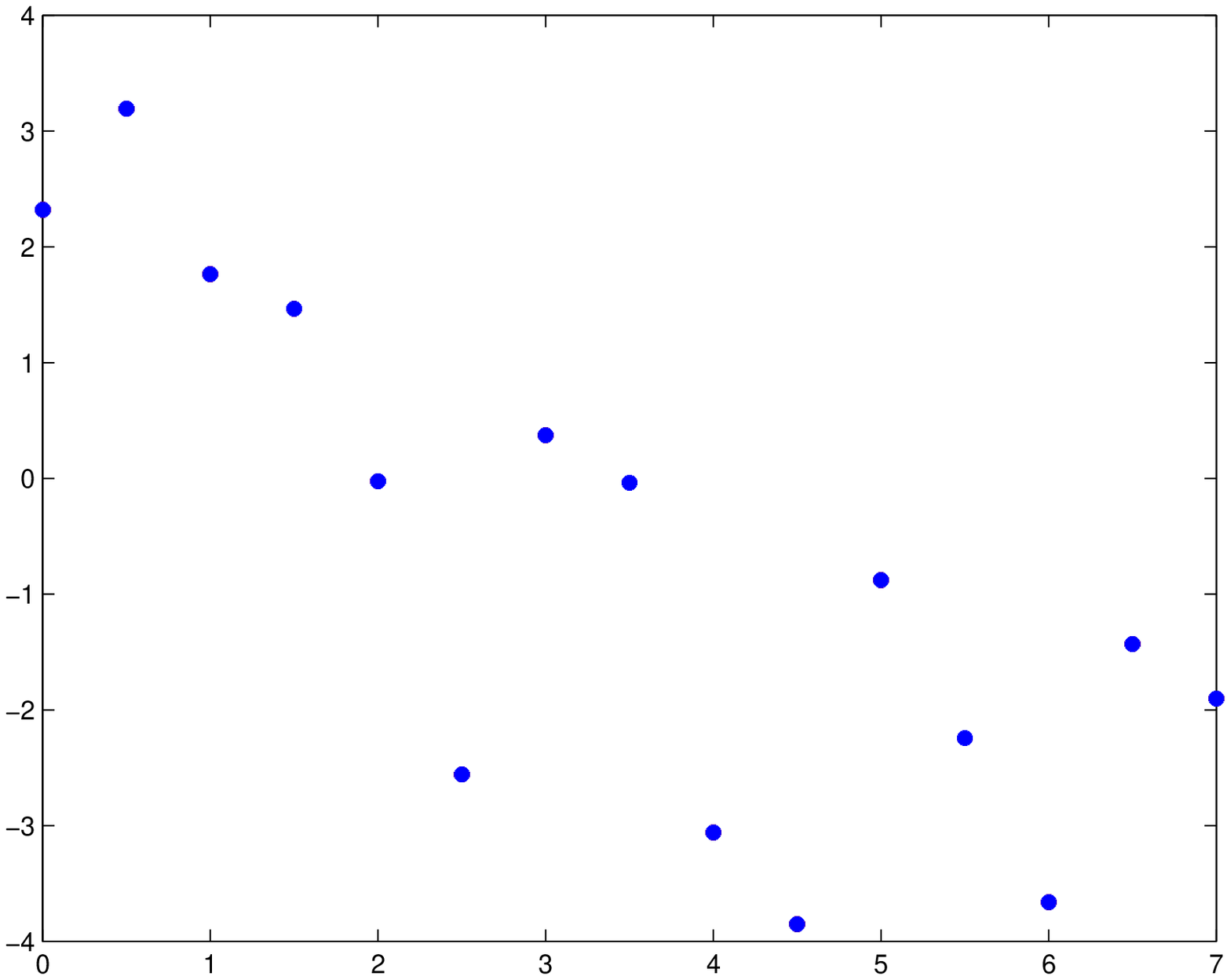} 
\end{tabular} 
&
\begin{tabular}[b]{@{}l@{}} 
$\disp{\wh{F}(q_i)=\sum_{j=1}^k A_{ij} \, \wh{x}_j}$ \\ 
and \\  
$\disp{F(q_i)=\sum_{j=1}^k A_{ij} \, x_j}$ 
\\ ~ \\ ~ \\  ~\\ ~\\ 
\end{tabular}
\\ 
Fig. 2:  
    a) $p(k,l|\yb)$, $p(l|\yb)$ and $p(k|\yb,\lh)$,\quad   
    b) original $\rho(r)$ \\ and estimated $\wh{\rho}(r)$,\quad    
    c) original $F(q_i)$ and estimated $\wh{F}(q_i)$.
\end{tabular}
	 
%\caption{a) The criterion $J(k,l)$,\quad b) estimated $\rho(r)$,\quad 
%c) estimated data.}
%\end{figure}

\newpage
%\begin{figure}
\begin{tabular}{@{}c@{}c@{}}
\epsfxsize=60mm\epsfysize=40mm\epsfbox{\FigDir simul1c.eps} & 
\epsfxsize=60mm\epsfysize=40mm\epsfbox{\FigDir 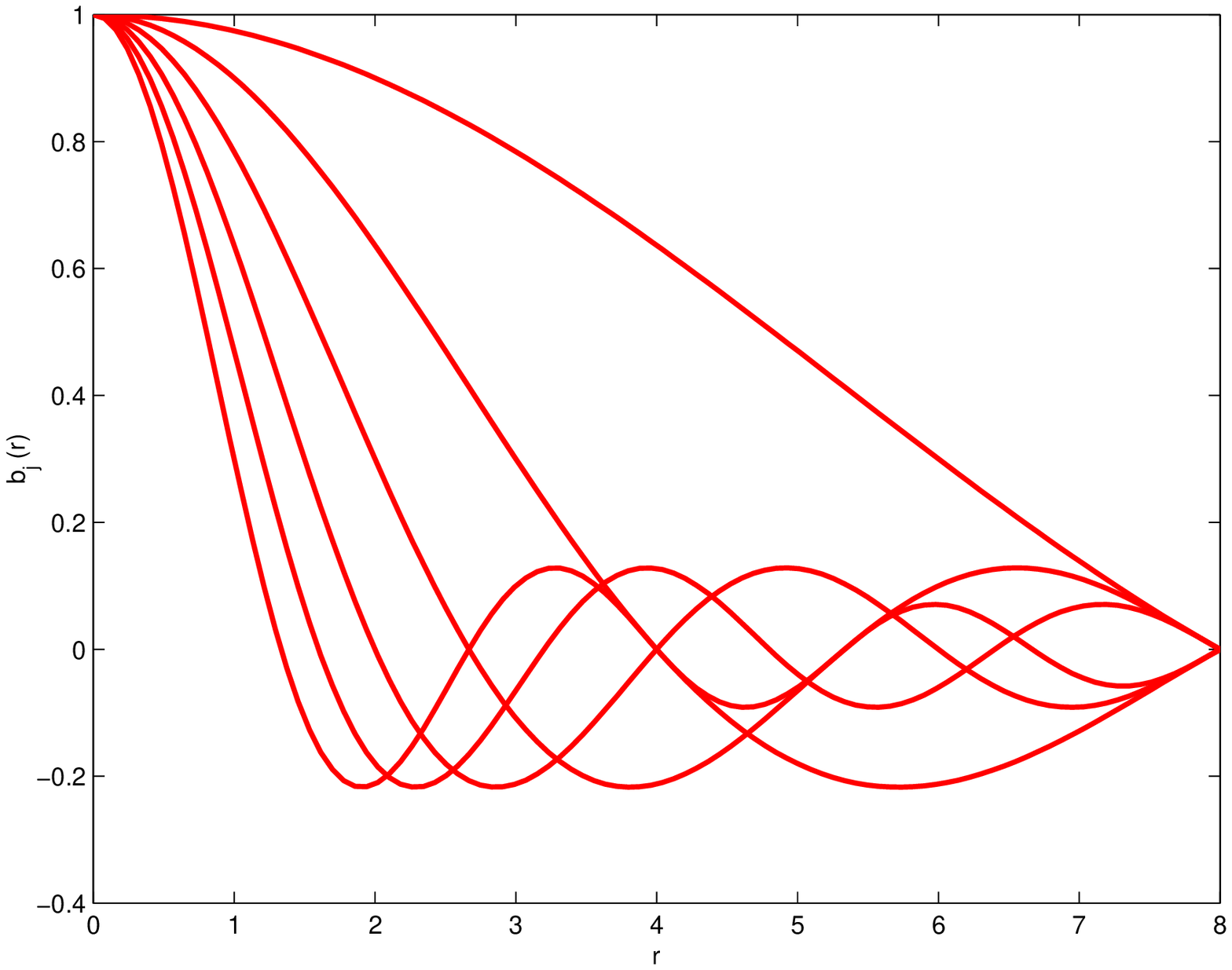} 
\\
\epsfxsize=60mm\epsfysize=40mm\epsfbox{\FigDir 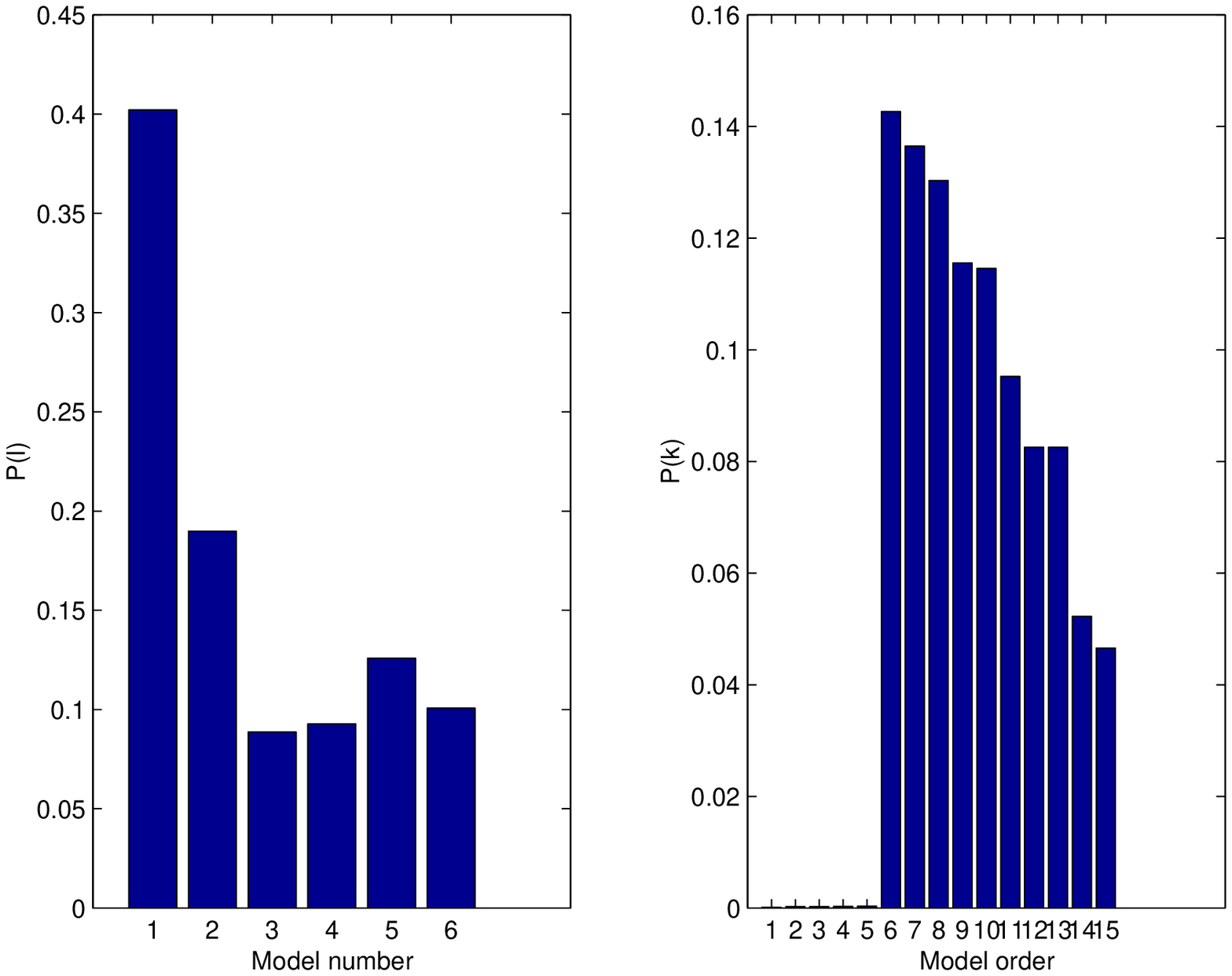} &
\epsfxsize=60mm\epsfysize=40mm\epsfbox{\FigDir 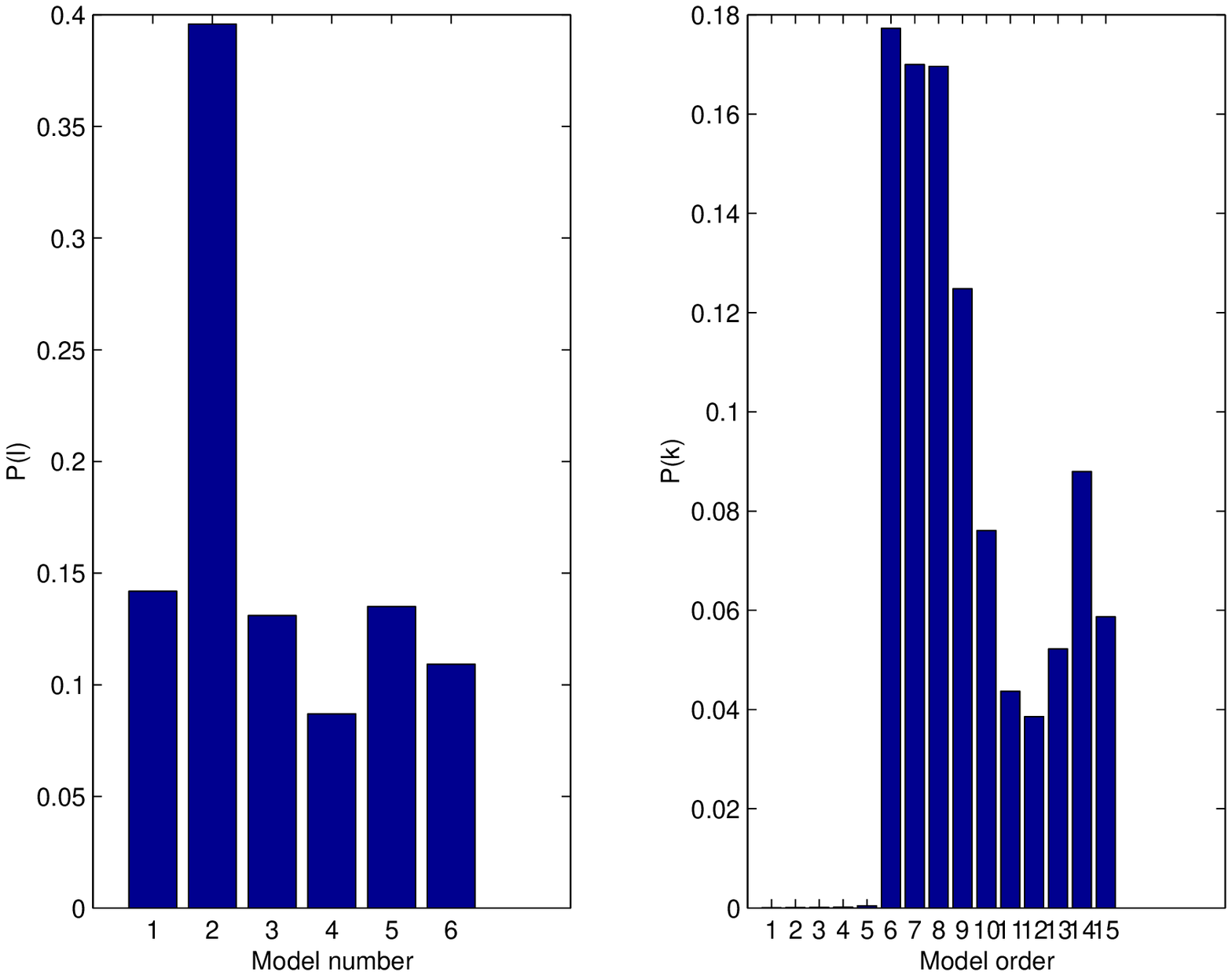} 
\\
\epsfxsize=60mm\epsfysize=40mm\epsfbox{\FigDir simul1d.eps} &
\epsfxsize=60mm\epsfysize=40mm\epsfbox{\FigDir 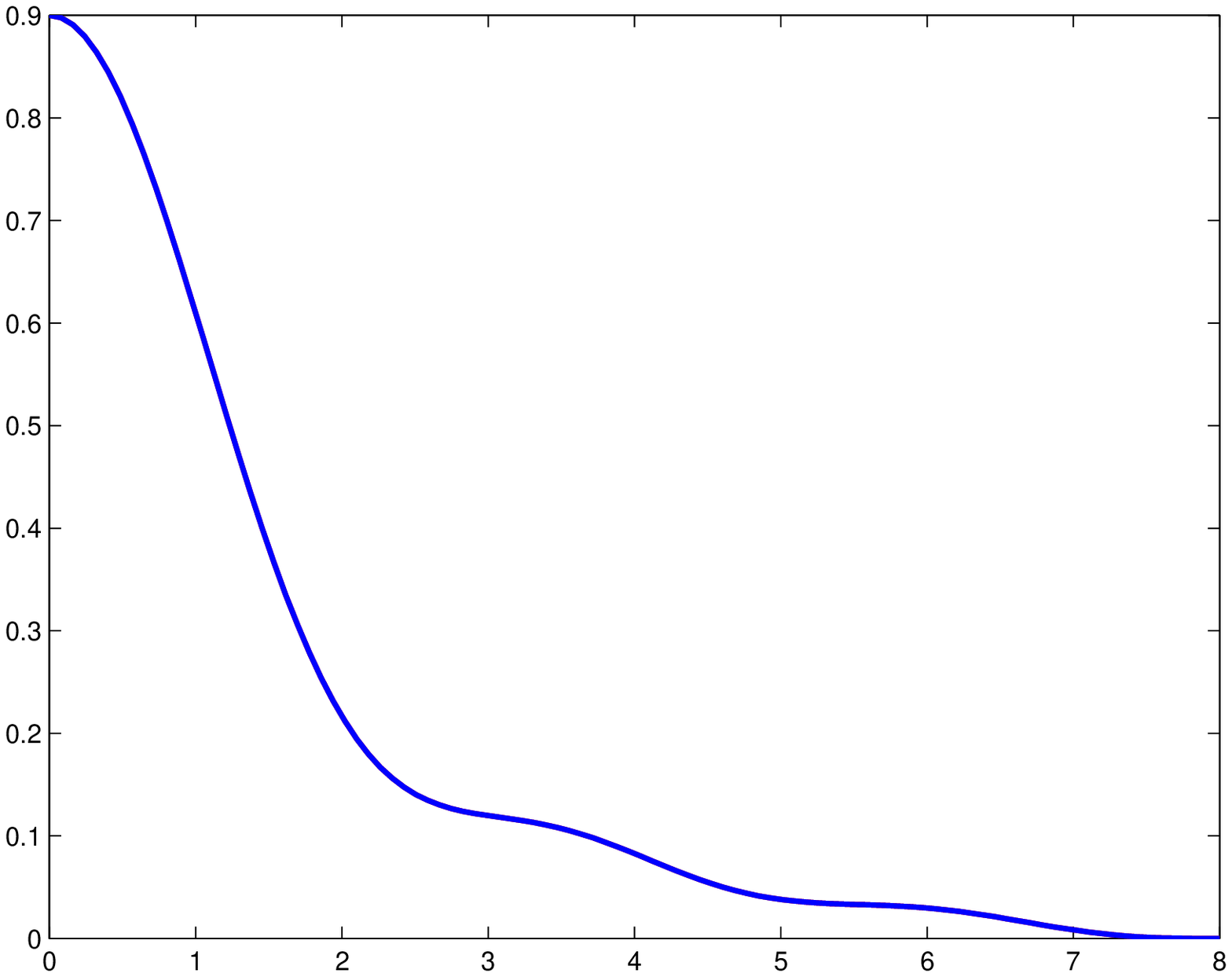} 
\\
\epsfxsize=60mm\epsfysize=40mm\epsfbox{\FigDir simul1e.eps} &
\epsfxsize=60mm\epsfysize=40mm\epsfbox{\FigDir 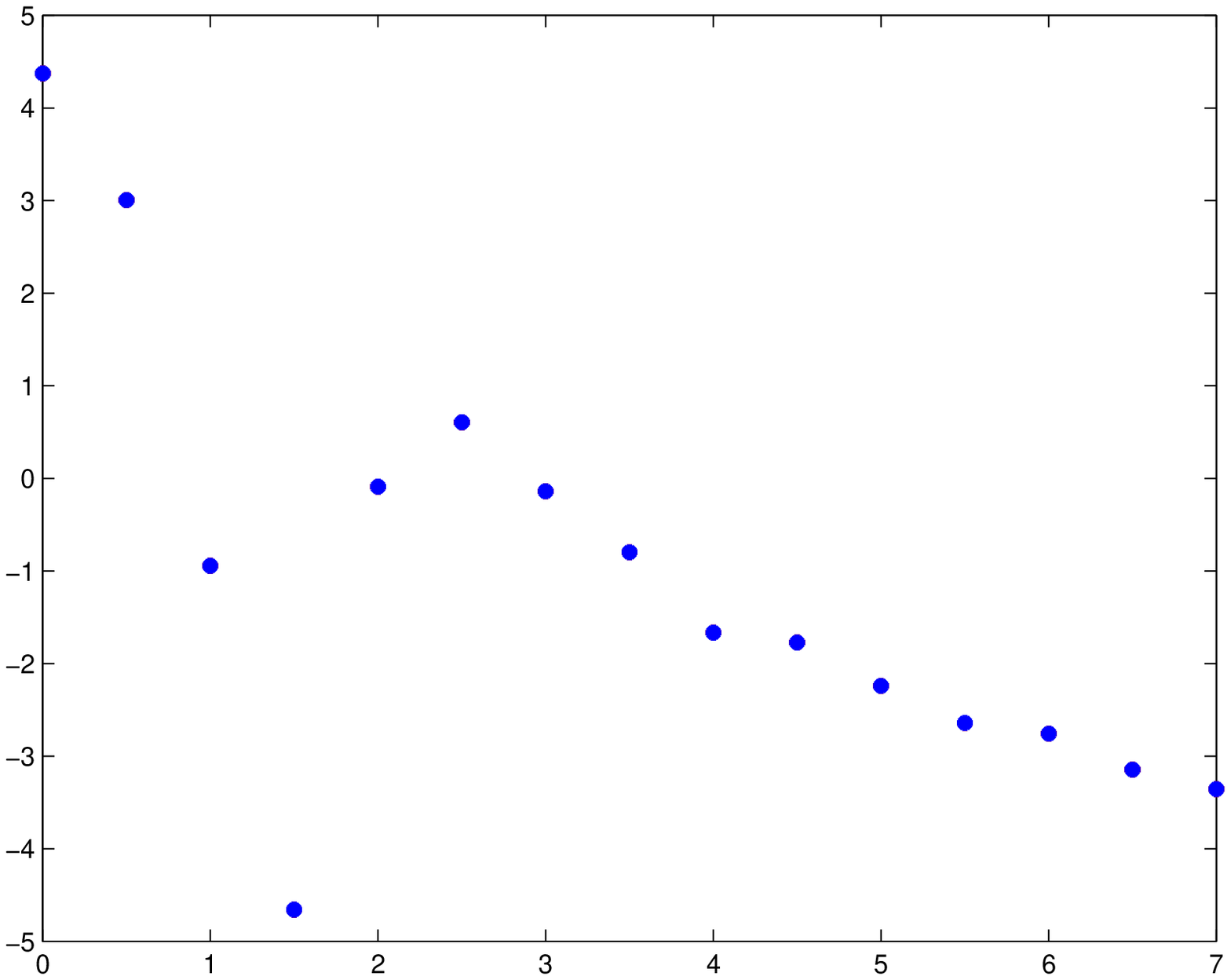} 
\end{tabular}
 
\bigskip\centerline{\btabu{lll}
 Fig. 3: & Left: $l=1$ \hspace*{2cm} & Right: $l=2$ \\ 
         & a) basis functions $b_j(r)$;  \\ 
			& b) $p(k,l|\yb)$;\\ 
			& c) $p(k|\yb)$ and $p(k|\yb,\lh)$; \\ 
			& d) $\rho(r)$ and $\wh{\rho}(r)$;\\  
			& e) $F(q_i)$ and $\wh{F}(q_i)$. 
\etabu}
%\end{figure}

\newpage
%\begin{figure}
\begin{tabular}{@{}c@{}c@{}}
\epsfxsize=60mm\epsfysize=40mm\epsfbox{\FigDir 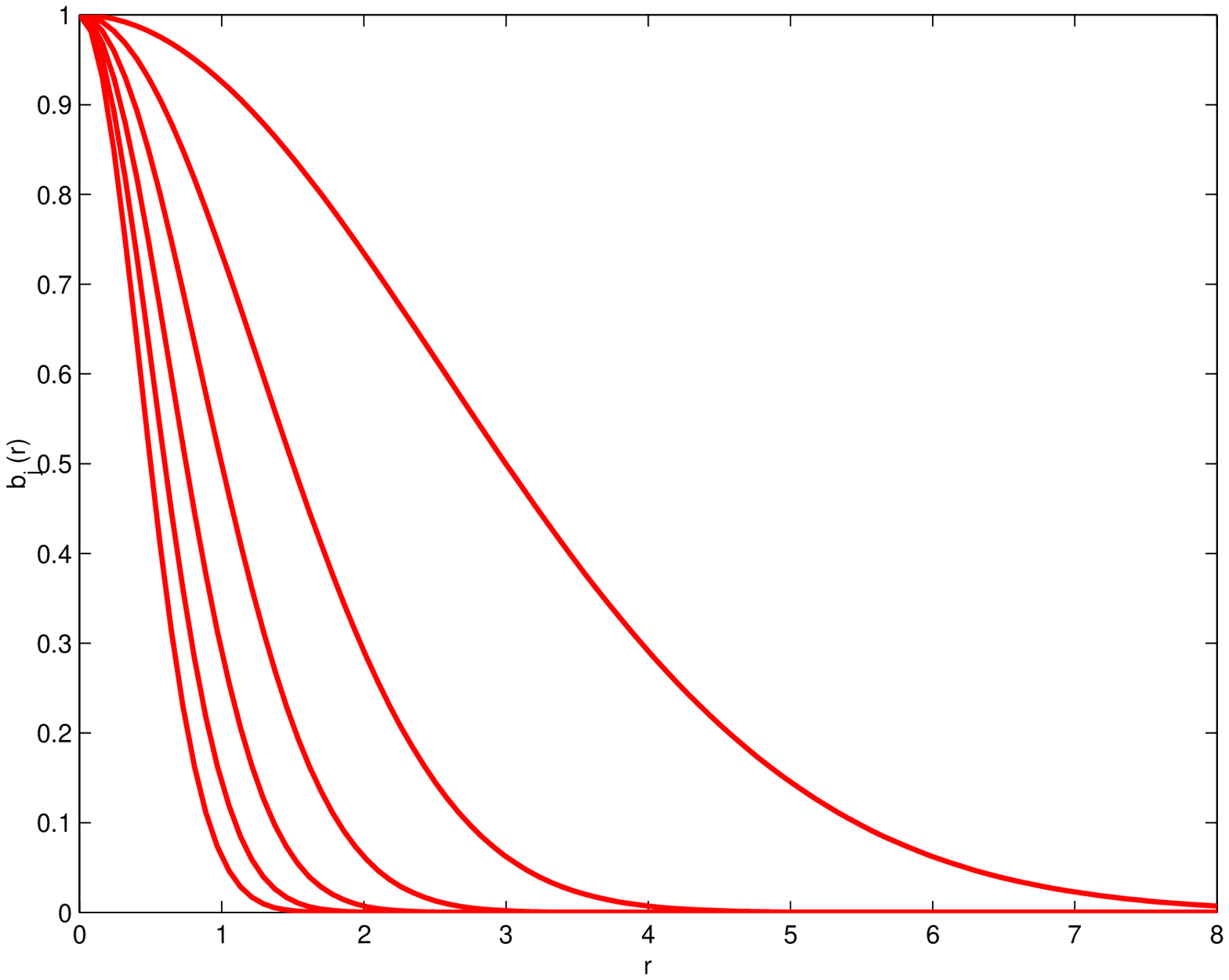} & 
\epsfxsize=60mm\epsfysize=40mm\epsfbox{\FigDir 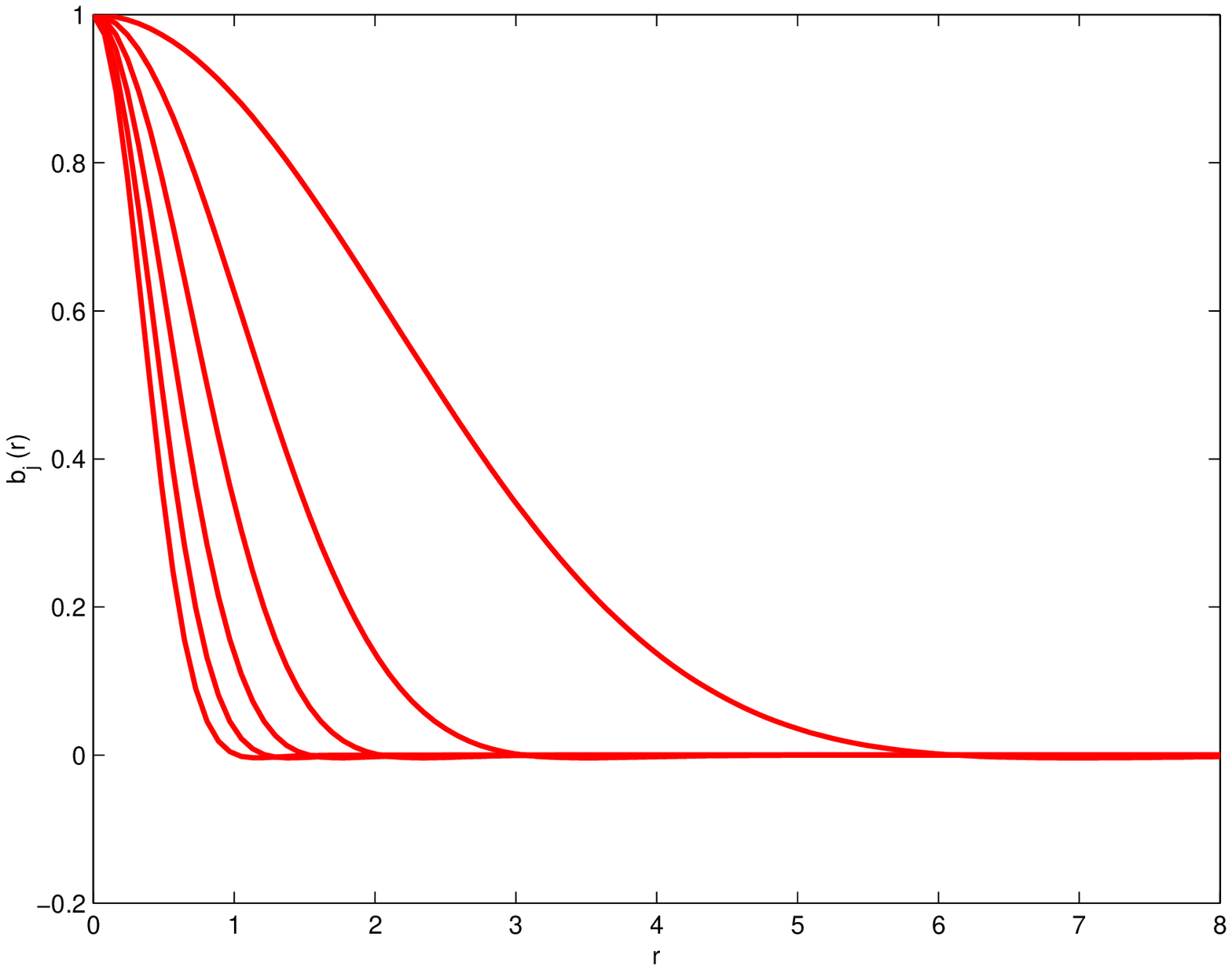} 
\\
\epsfxsize=60mm\epsfysize=40mm\epsfbox{\FigDir 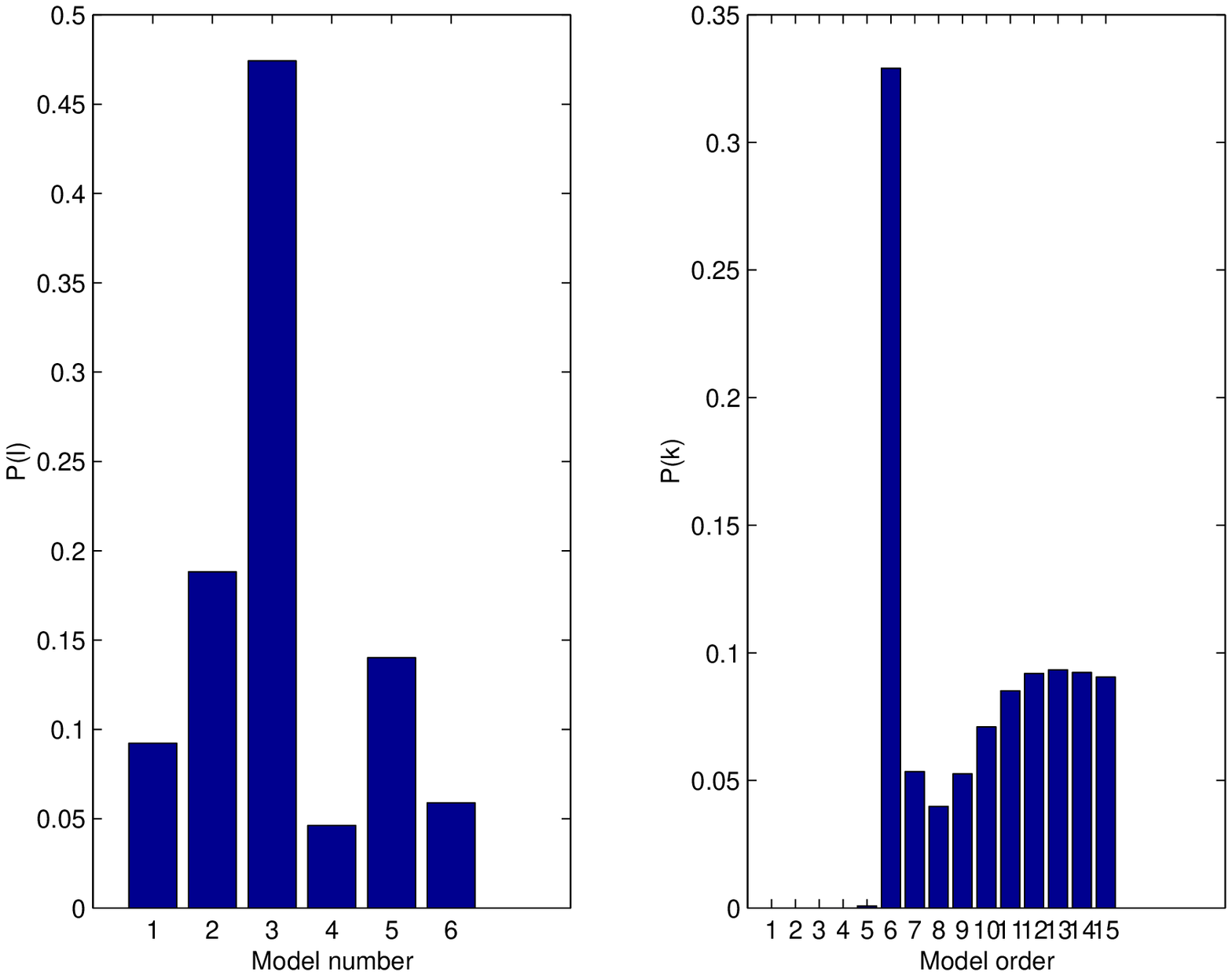} &
\epsfxsize=60mm\epsfysize=40mm\epsfbox{\FigDir 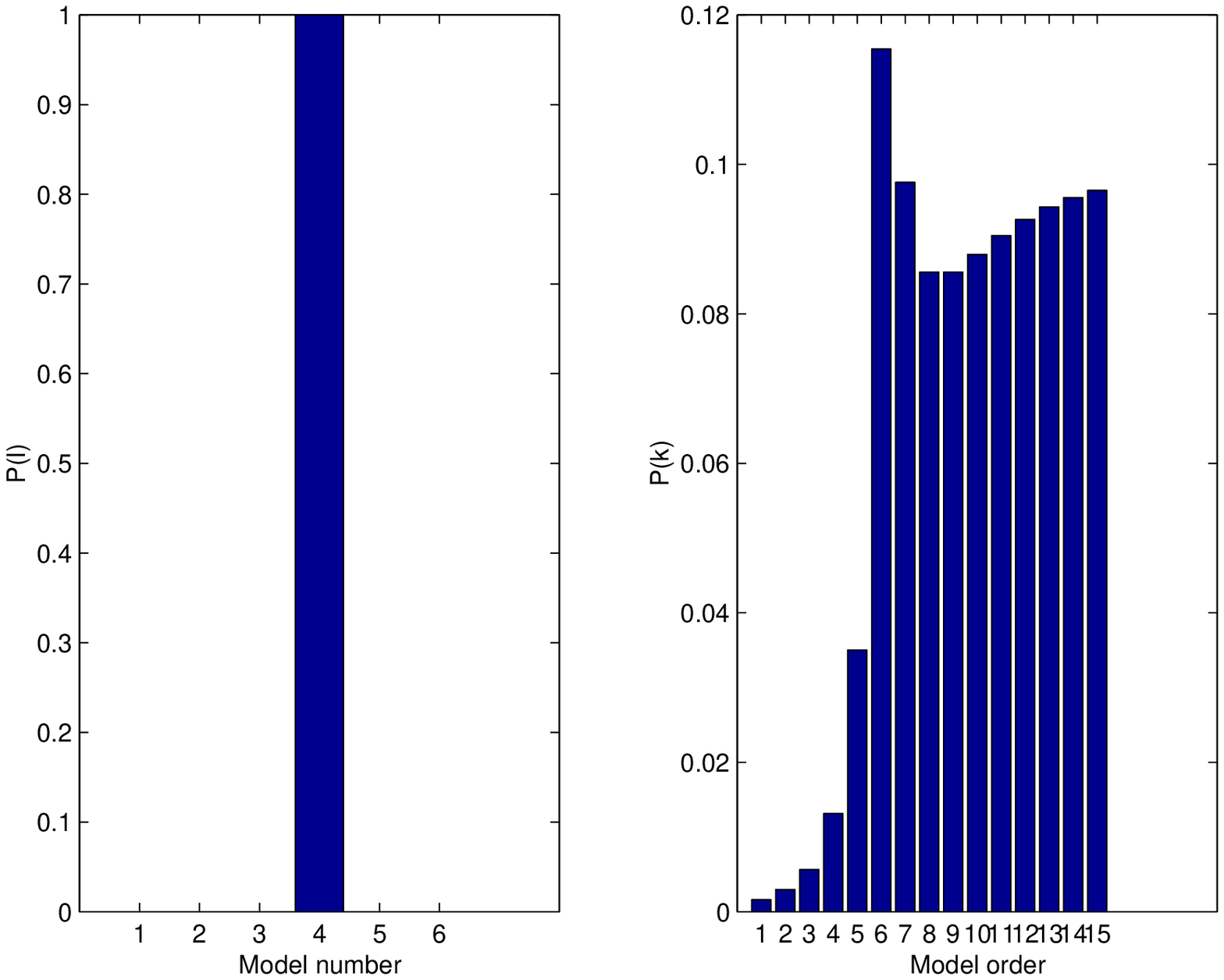} 
\\ 
\epsfxsize=60mm\epsfysize=40mm\epsfbox{\FigDir 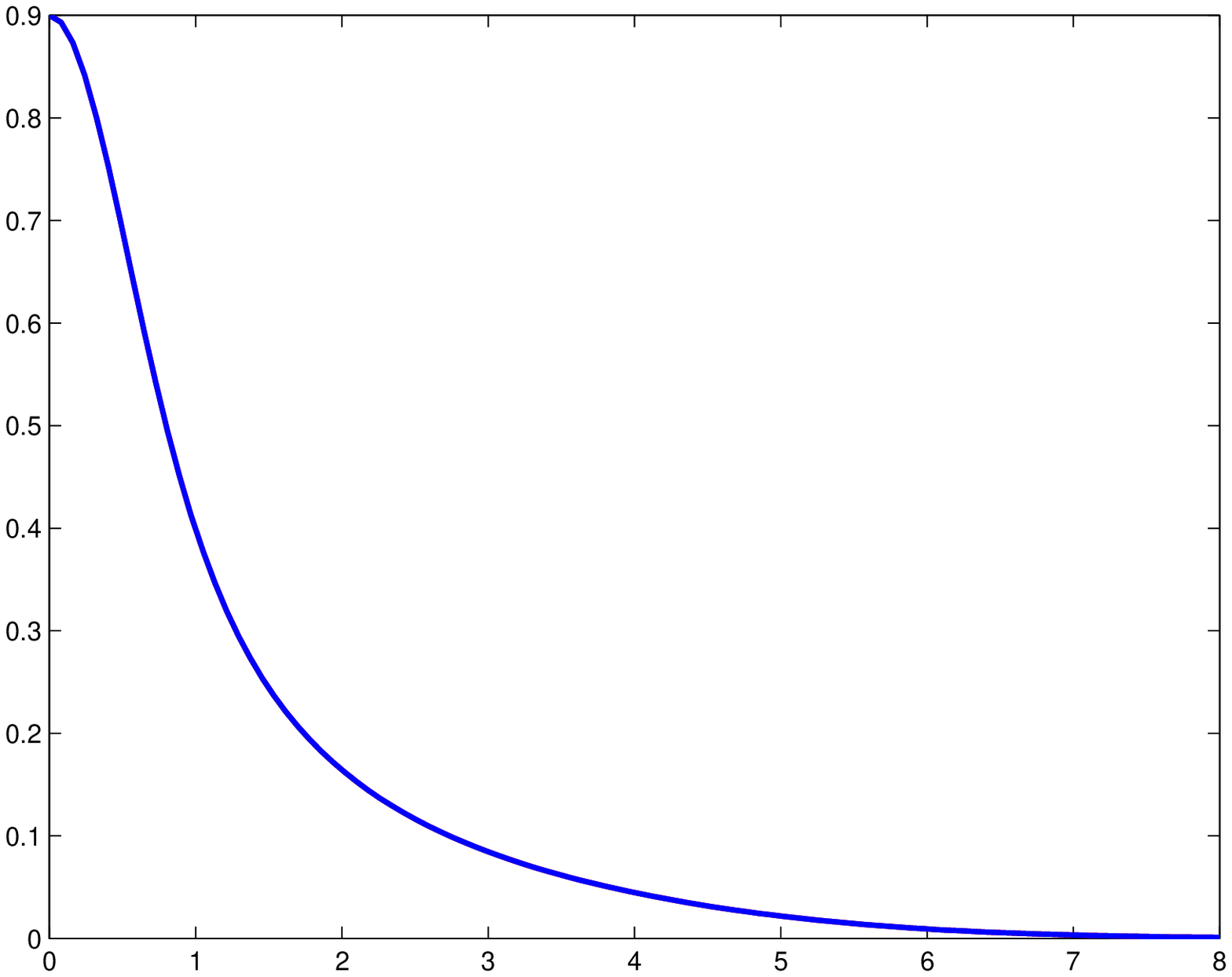} &
\epsfxsize=60mm\epsfysize=40mm\epsfbox{\FigDir 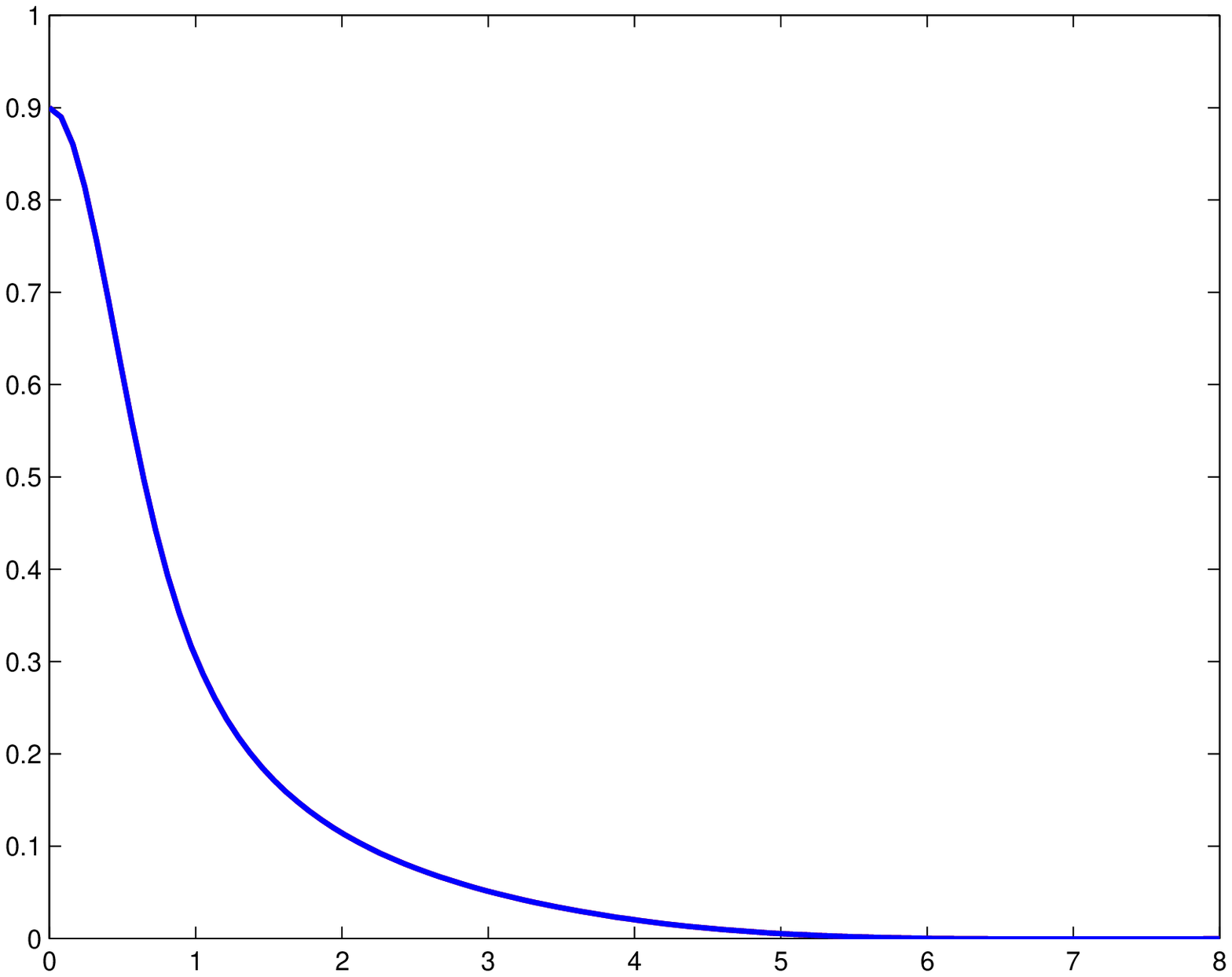} 
\\
\epsfxsize=60mm\epsfysize=40mm\epsfbox{\FigDir 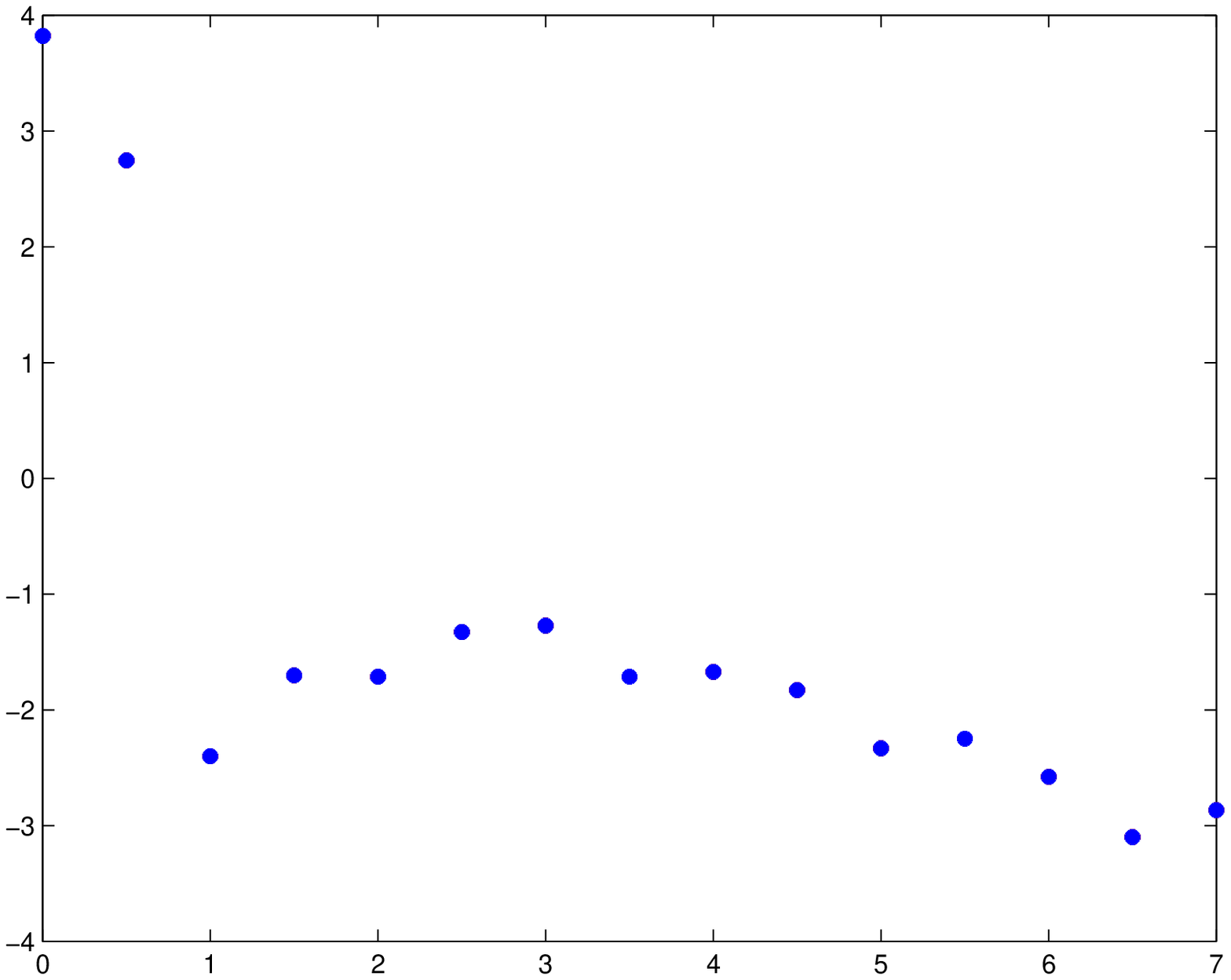} &
\epsfxsize=60mm\epsfysize=40mm\epsfbox{\FigDir 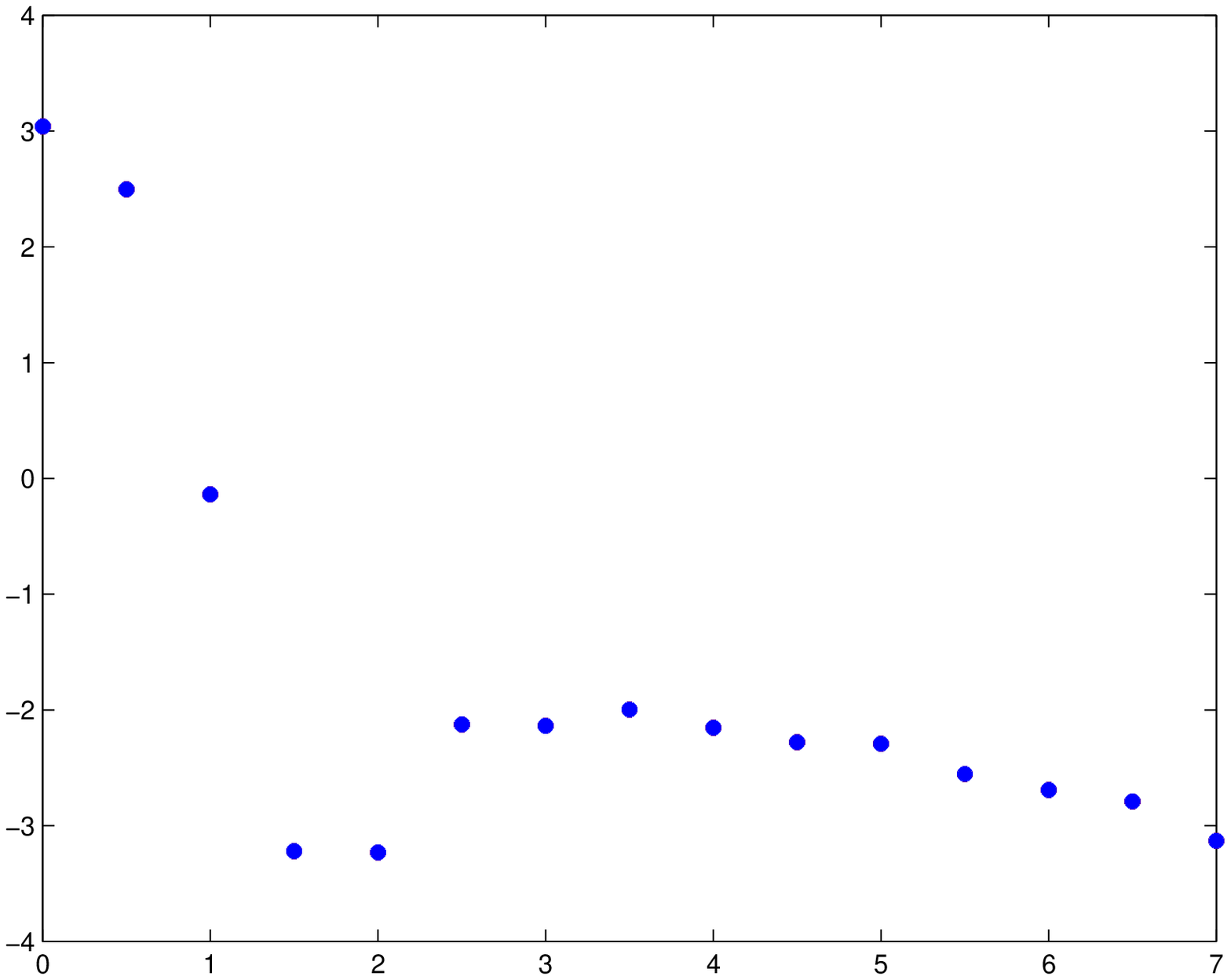} 
\end{tabular}
 
\bigskip\centerline{\btabu{lll}
 Fig. 4: & Left: $l=3$ \hspace*{2cm} & Right: $l=4$ \\ 
         & a) basis functions $b_j(r)$;  \\ 
			& b) $p(k,l|\yb)$;\\ 
			& c) $p(k|\yb)$ and $p(k|\yb,\lh)$; \\ 
			& d) $\rho(r)$ and $\wh{\rho}(r)$;\\  
			& e) $F(q_i)$ and $\wh{F}(q_i)$. 
\etabu}
%\end{figure}

\newpage
%\begin{figure}
\begin{tabular}{@{}c@{}c@{}}
\epsfxsize=60mm\epsfysize=40mm\epsfbox{\FigDir 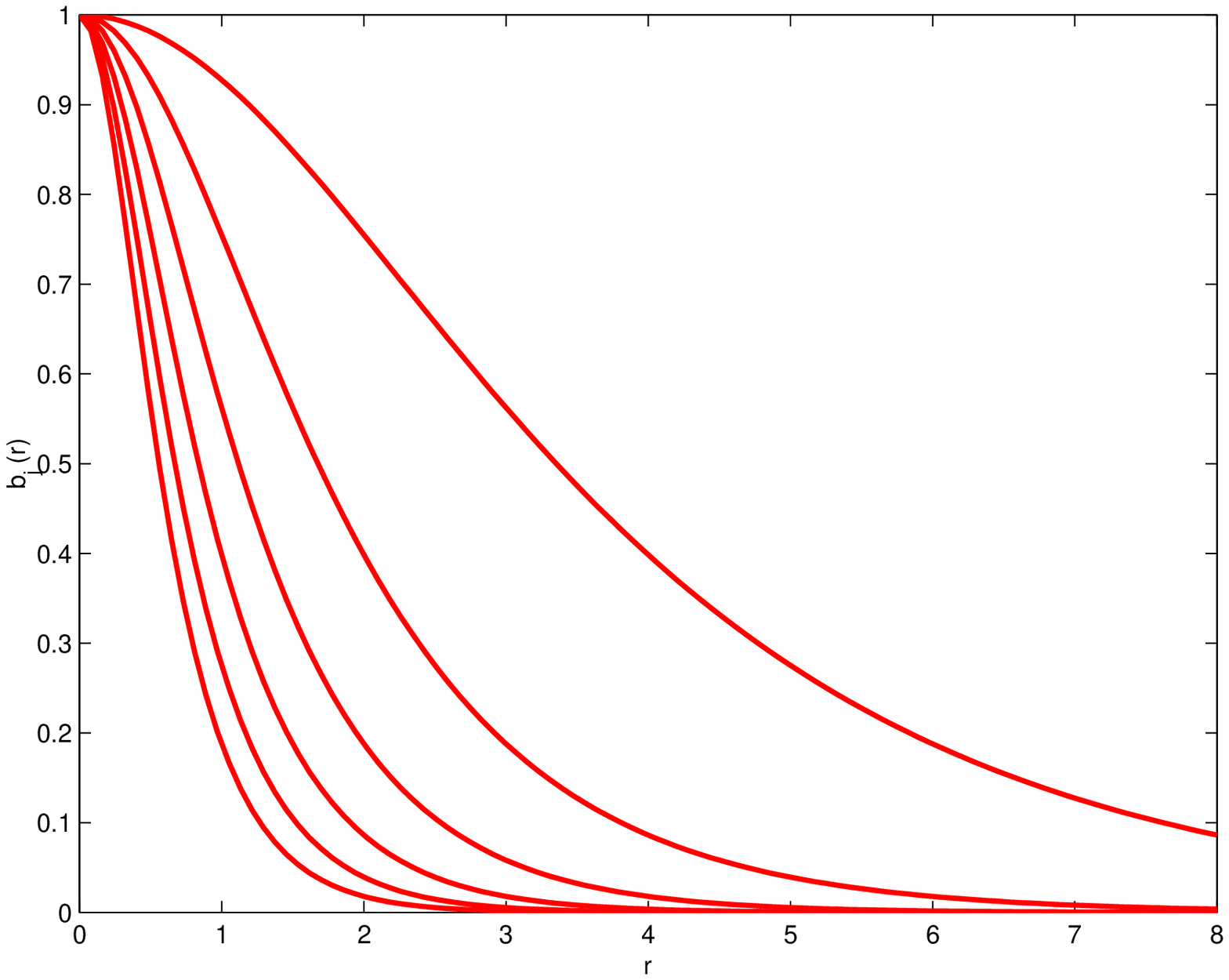} & 
\epsfxsize=60mm\epsfysize=40mm\epsfbox{\FigDir 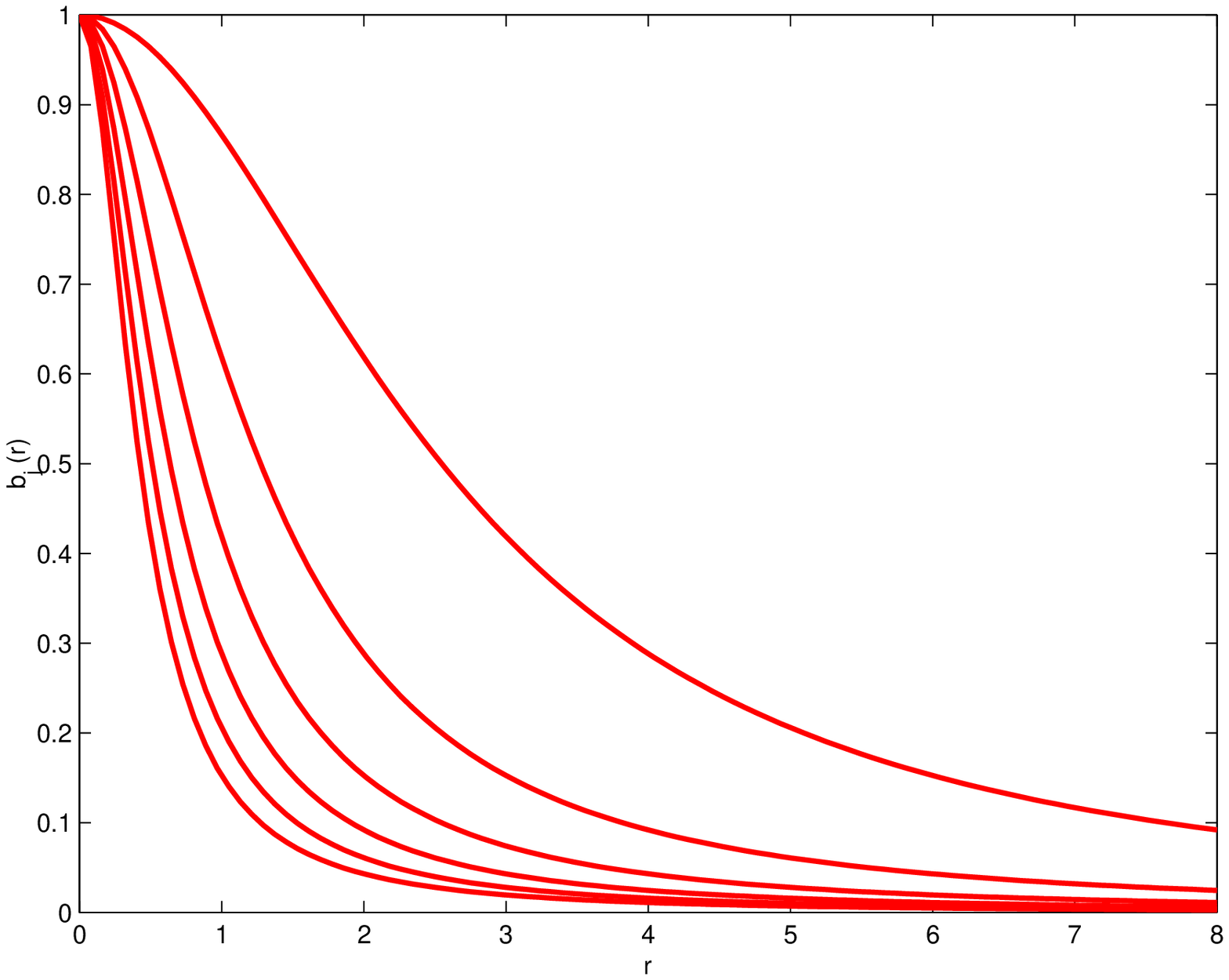} 
\\
\epsfxsize=60mm\epsfysize=40mm\epsfbox{\FigDir 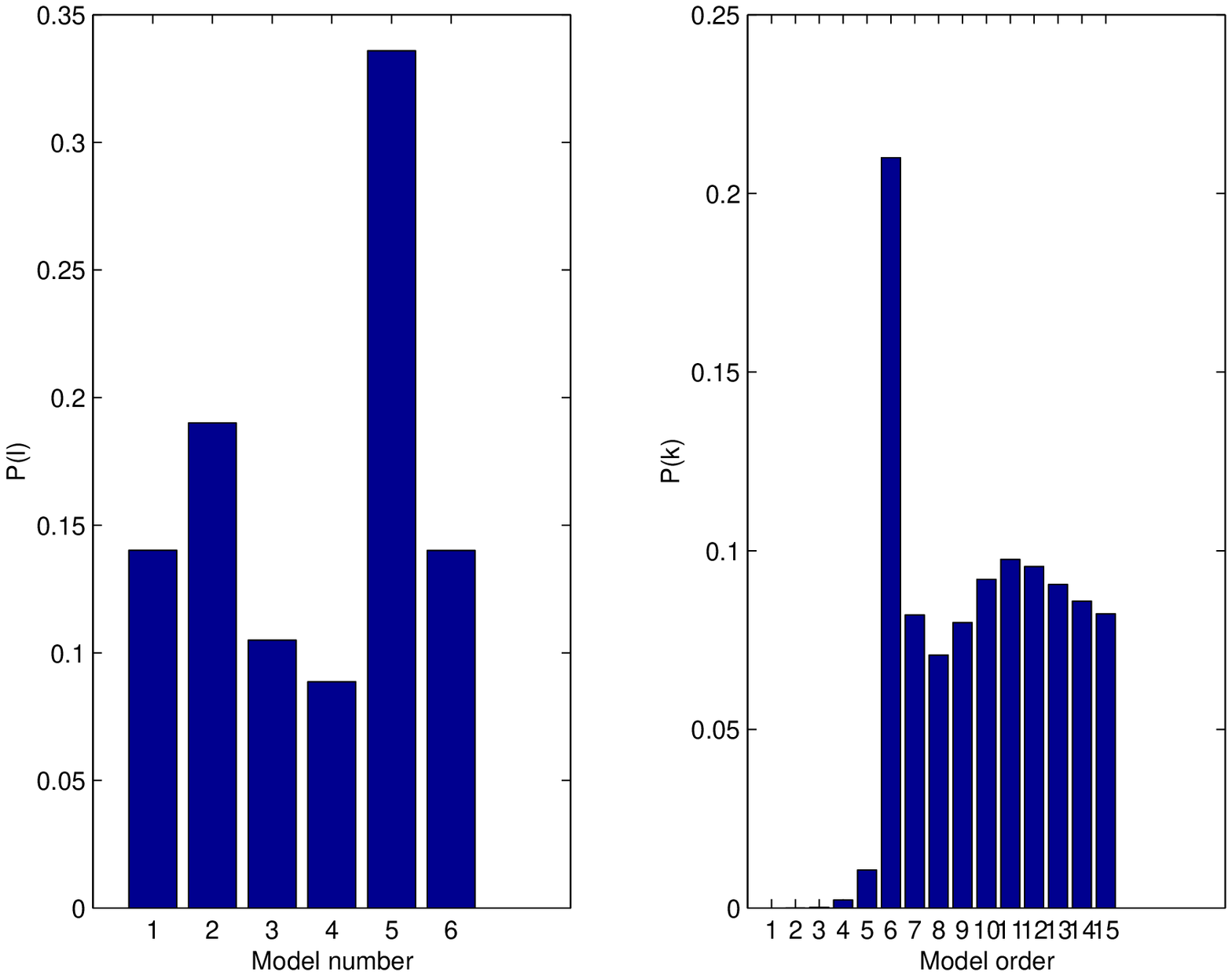} &
\epsfxsize=60mm\epsfysize=40mm\epsfbox{\FigDir 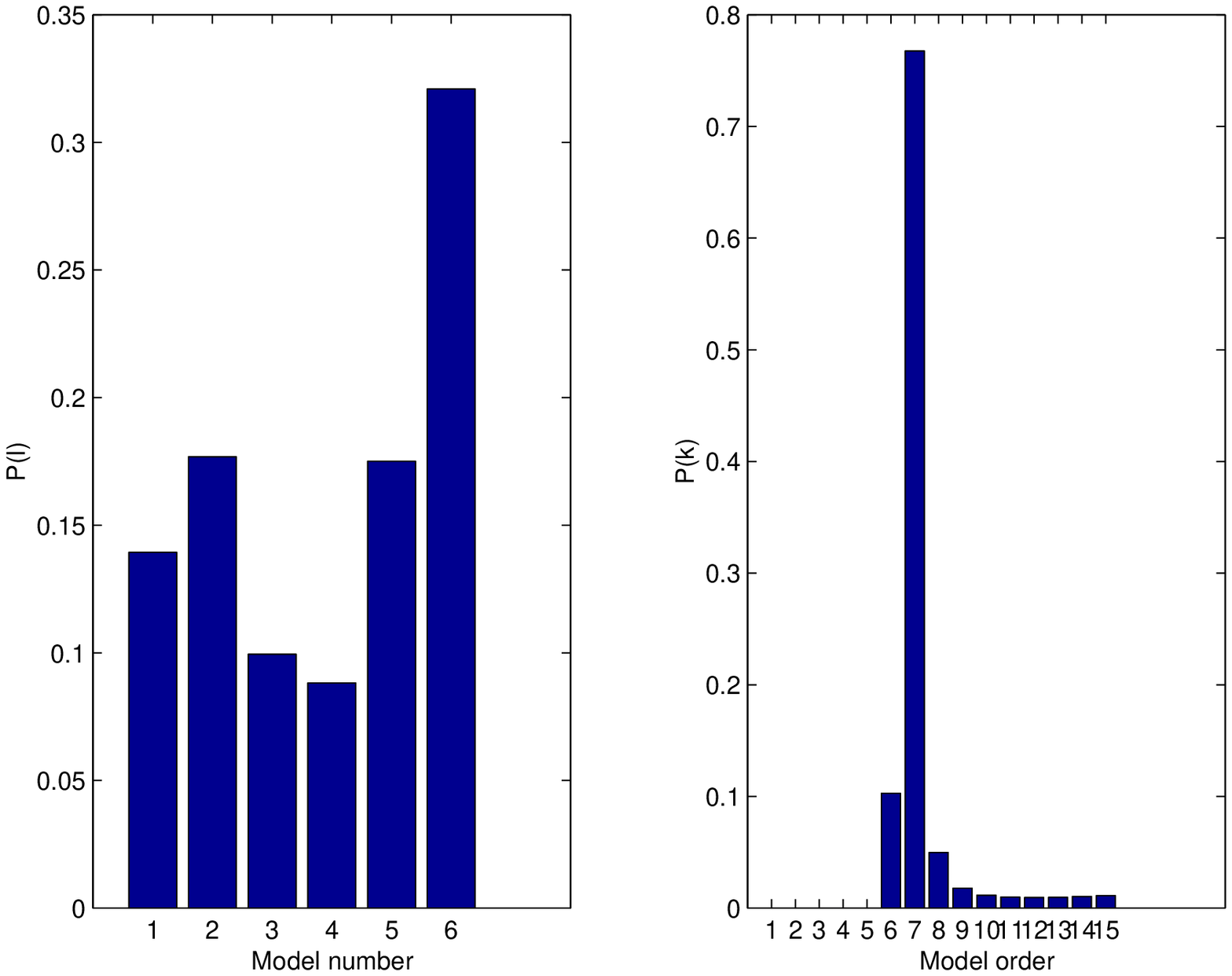} 
\\ 
\epsfxsize=60mm\epsfysize=40mm\epsfbox{\FigDir 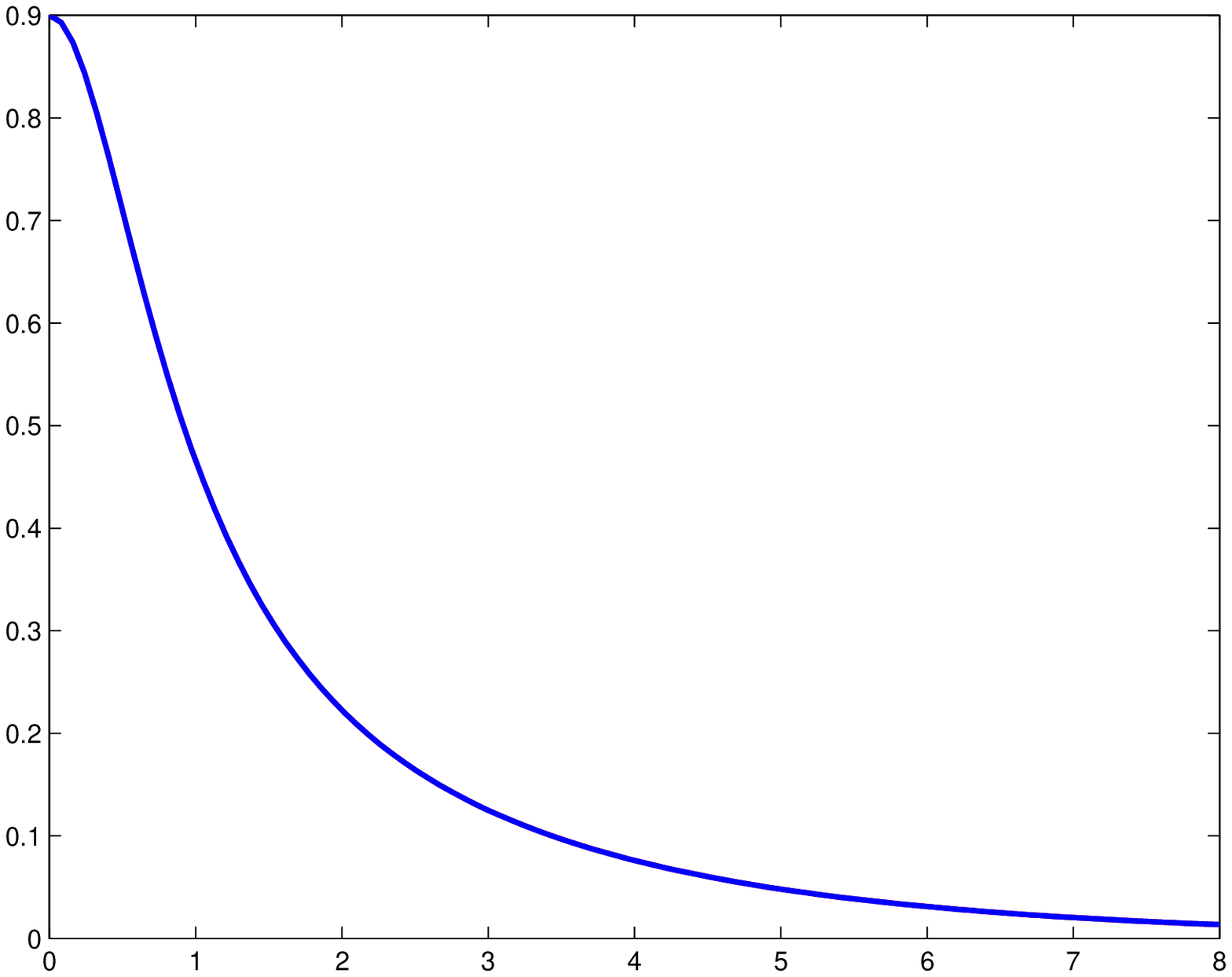} &
\epsfxsize=60mm\epsfysize=40mm\epsfbox{\FigDir 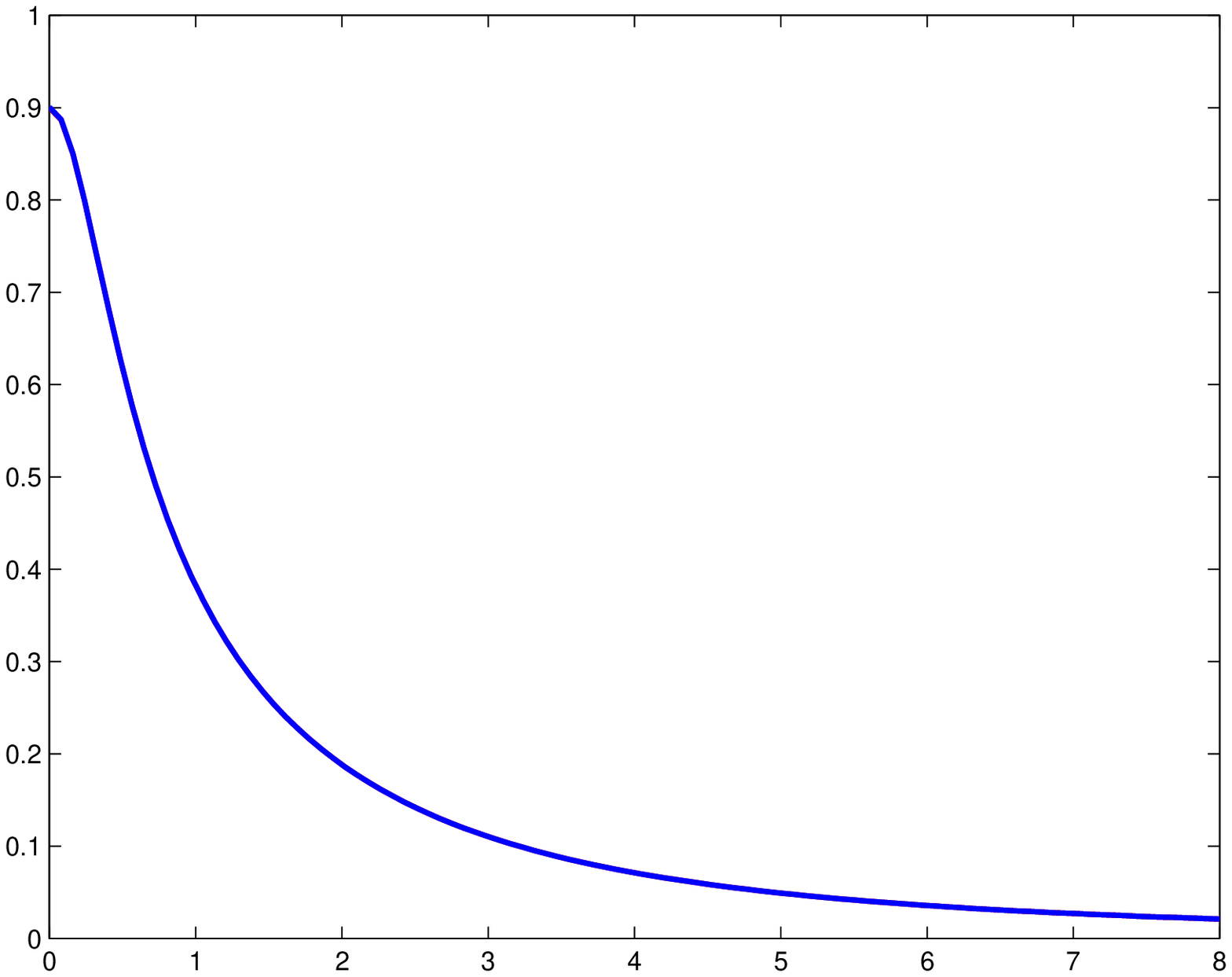} 
\\
\epsfxsize=60mm\epsfysize=40mm\epsfbox{\FigDir 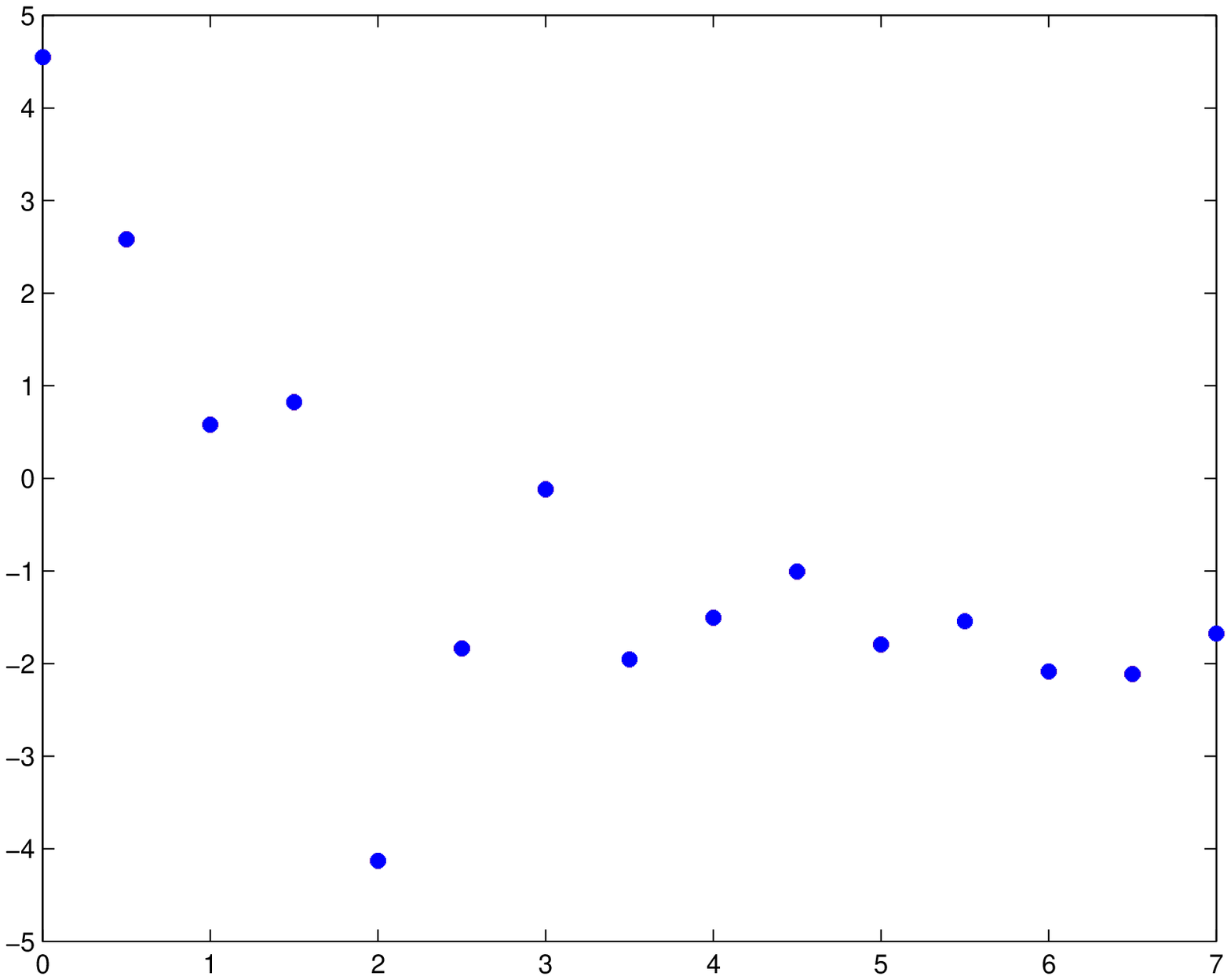} &
\epsfxsize=60mm\epsfysize=40mm\epsfbox{\FigDir 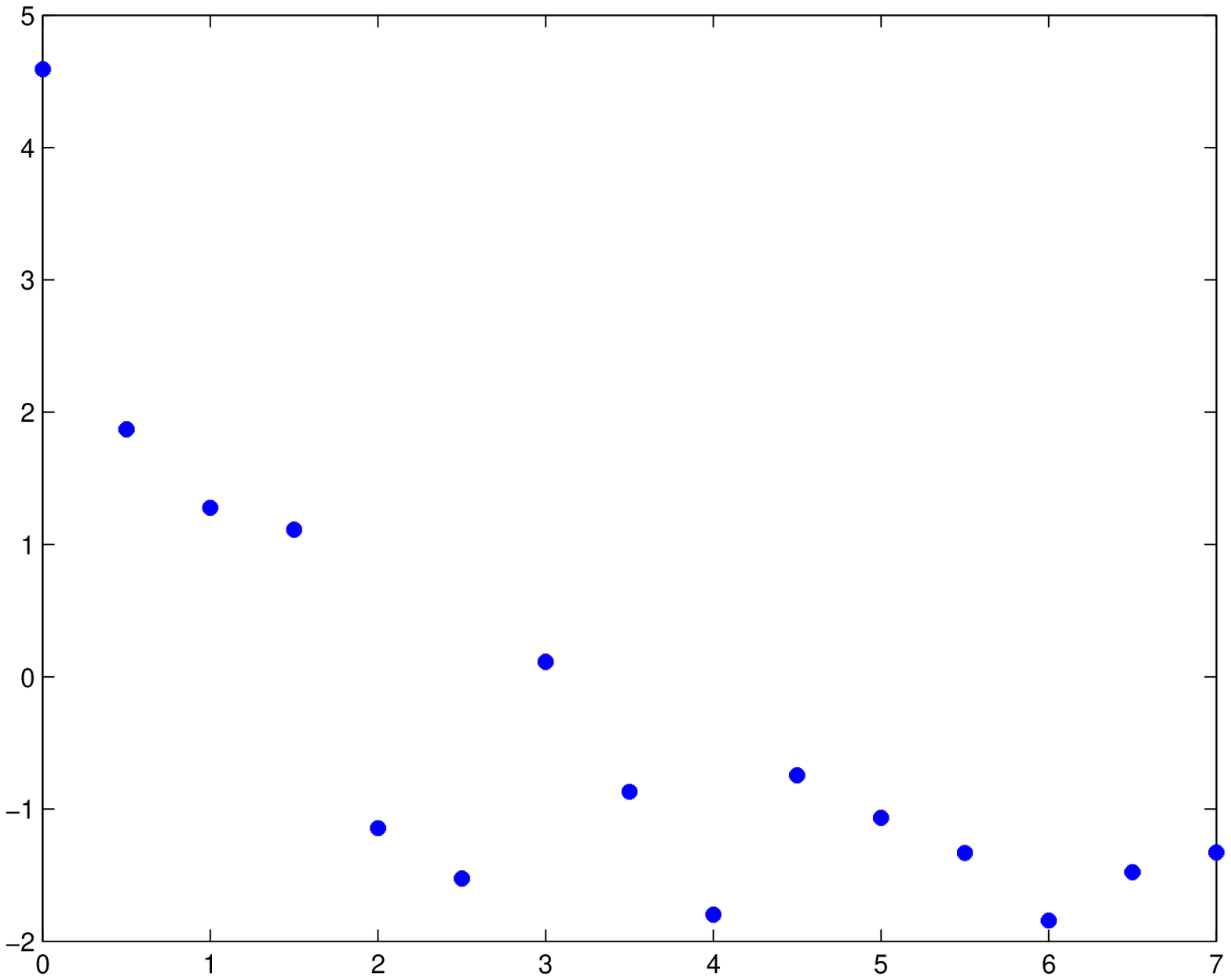} 
\end{tabular}
\bigskip\centerline{\btabu{lll}
 Fig. 5: & Left: $l=5$ \hspace*{2cm} & Right: $l=6$ \\ 
         & a) basis functions $b_j(r)$;  \\ 
			& b) $p(k,l|\yb)$;\\ 
			& c) $p(k|\yb)$ and $p(k|\yb,\lh)$; \\ 
			& d) $\rho(r)$ and $\wh{\rho}(r)$;\\  
			& e) $F(q_i)$ and $\wh{F}(q_i)$. 
\etabu}
%\end{figure}

\newpage
Note that in these tests, we know perfectly the model and generated the data 
according to our hypothesis. To test the method to a more realistic case, 
we chose a model for which we can have an exact analytic expression for the 
integrals. For example, if we choose a symmetric Fermi 
distribution~\cite{GLMP91}
\begin{equation}\label{Trho}
\rho(r) =\alpha\,\frac{\cosh(R/d)}{\cosh(R/d)+\cosh(r/d)}
\end{equation}
an analytical expression for the corresponding charge form factor
can easily be obtained~\cite{K91}:
\begin{equation}\label{TFc}
F(q) = -\frac{4\pi^2\alpha d}{q} \, \frac{\cosh(R/d)}{\sinh(R/d)}
\, \left[\frac{R\,\cos(q R)}{\sinh(\pi q d)} -\frac{\pi d\sin(q R)
\cosh(\pi q d)}{\sinh^2(\pi q d)}\right].
\end{equation}
Only two of the parameters $\alpha$, $R$ and $d$ are independent
since the charge density must fulfill the normalization condition
\begin{equation}\label{Norm}
4\pi\int r^2\,\rho(r)\d{r} = Z.
\end{equation}

Figure~6 shows the theoretical charge density $\rho(r)$
of ~$^{12}$C ~(Z=6) obtained from (\ref{Trho})
for $r\in[0, 8]$ fm with $R=1.1$ $A$ and $d=0.626$ fm
and the theoretical charge form factor $F(q)$ obtained by
(\ref{TFc}) for $q\in[0,8]$ fm$^{-1}$ and the 15 simulated 
data:
$$
\bm{q}=[0.001, .5, 1.0, 1.5, 2.0, 2.5, 3.0, 3.5, 4.0, 4.5, 5.0, 5.5, 6.0, 
6.5, 7.0] \, \mbox{fm}^{-1}
$$
\rem{$$
\bm{q}=[0.001, .5, 1.0, 2.0, 3.0, 4.0, 5.0, 6.0, 7.0] \, \mbox{fm}^{-1}
$$
}
which are used as inputs to the inversion method. 

\bcc
\begin{tabular}{@{}cc@{}}
\epsfxsize=60mm\epsfysize=60mm
\epsfbox{\FigDir 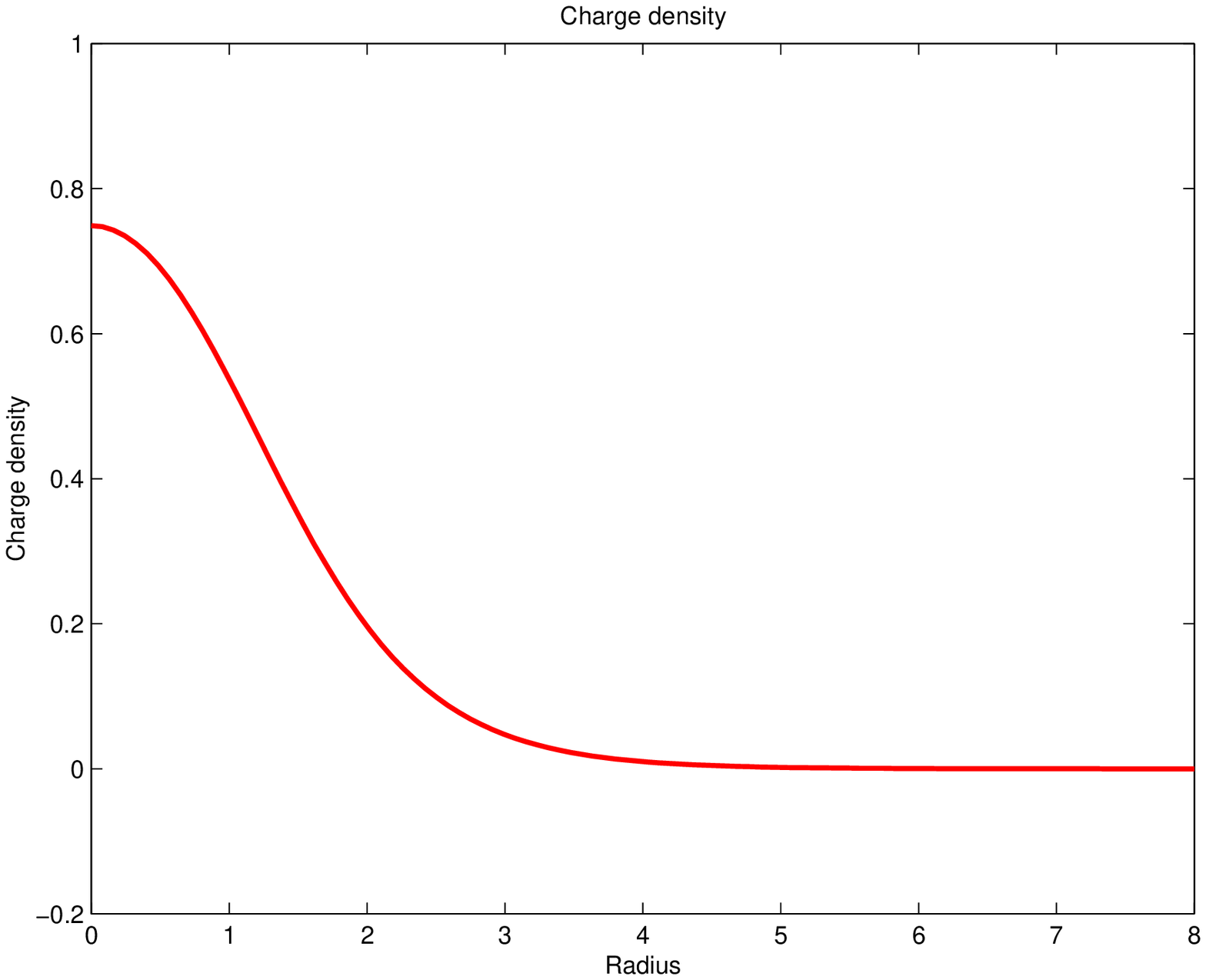} &
\epsfxsize=60mm\epsfysize=60mm
\epsfbox{\FigDir 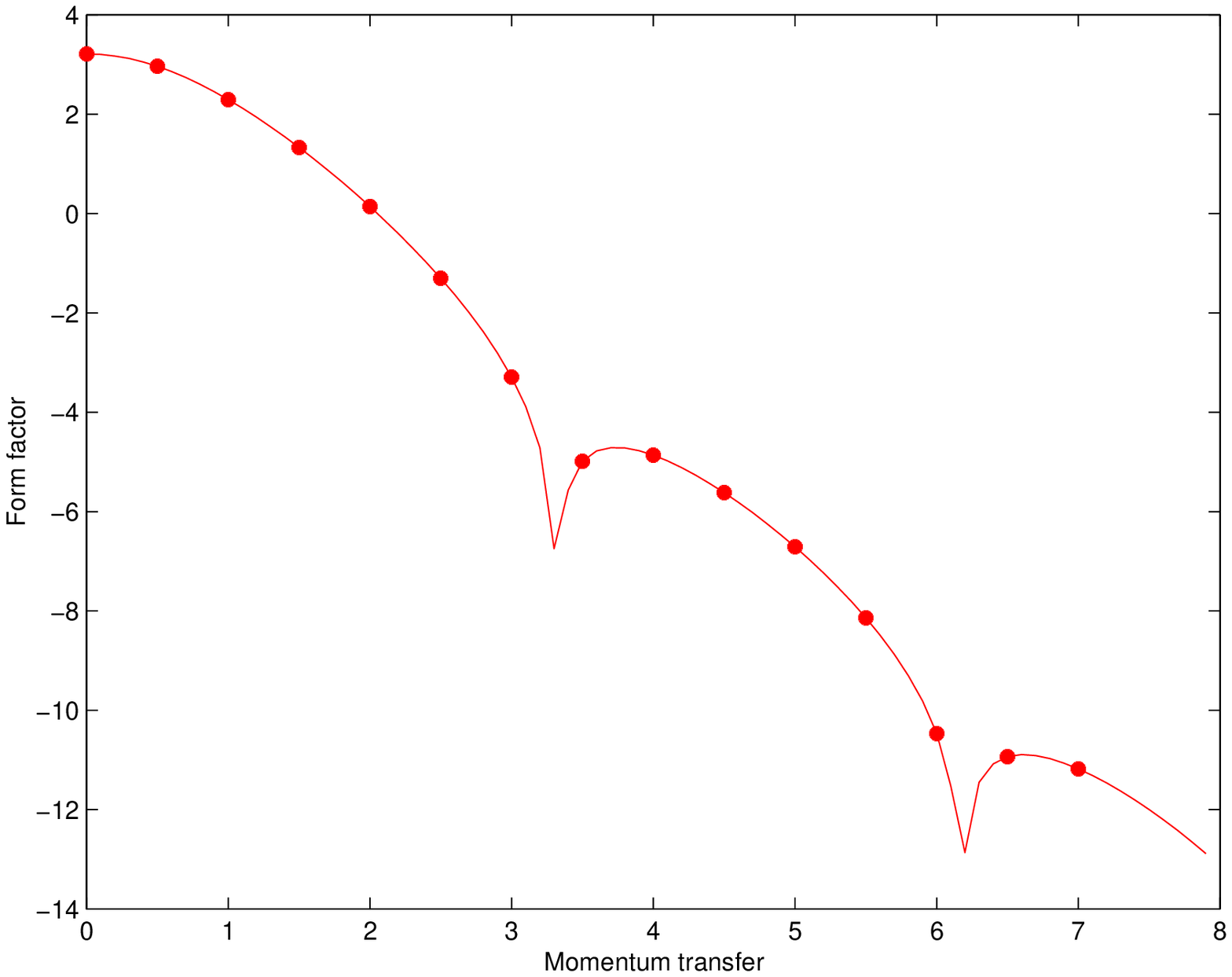} 
\end{tabular}
\\ 
{Fig. 6: Theoretical charge density $\rho(r)$,
charge form factor $\log| F(q) |$ and the data
[stars] used for numerical experiments [right].}
\ecc

First note that, even with the exact data, there are an infinite number 
of solutions which fits exactly the data. The following figure shows a 
few set of these solutions.  

\bcc
\begin{tabular}{@{}c@{}}
\epsfxsize=120mm\epsfysize=80mm\epsfbox{\FigDir 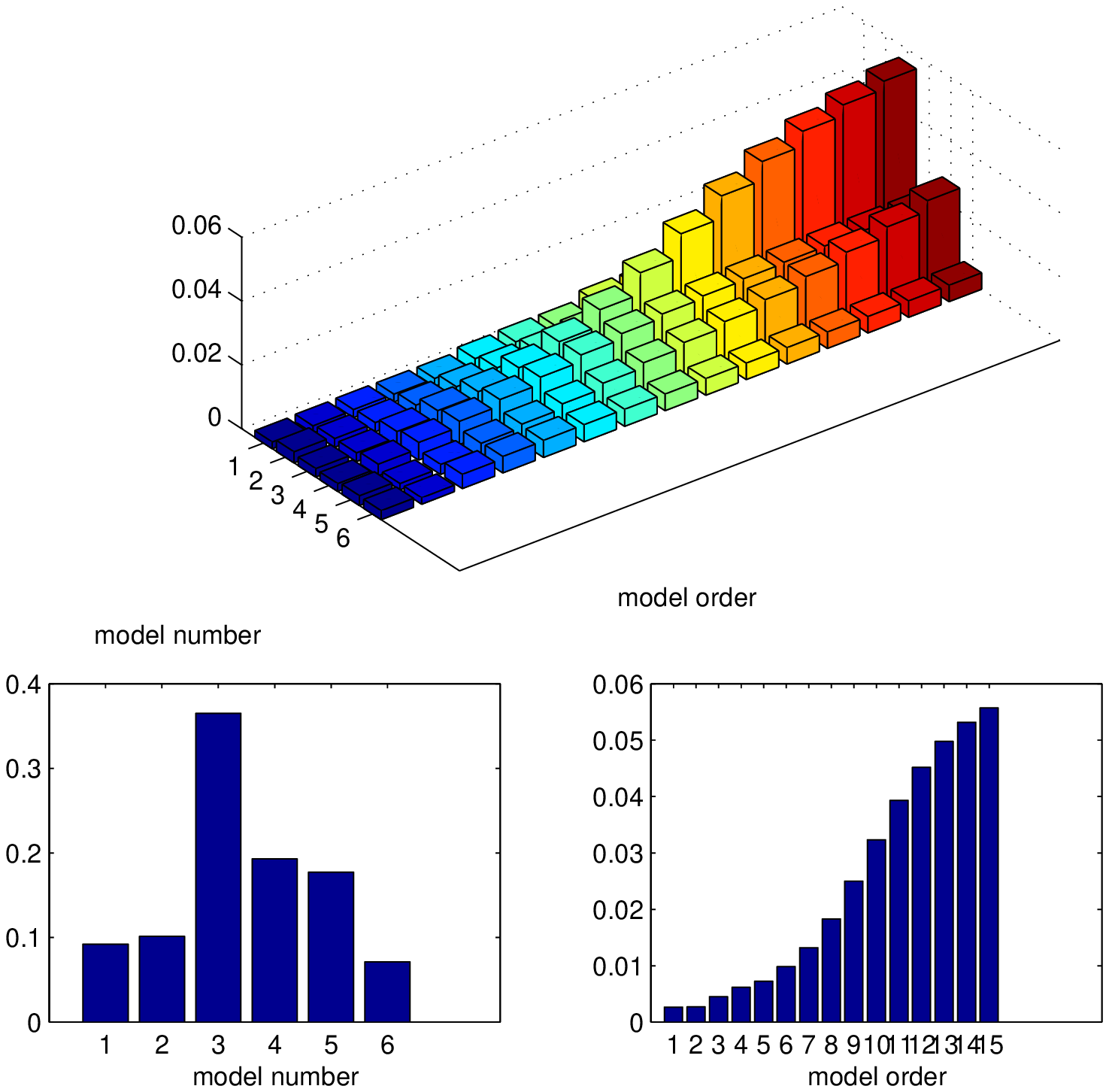} 
\\ 
\begin{tabular}{@{}c@{}c@{}}
\epsfxsize=60mm\epsfysize=60mm\epsfbox{\FigDir 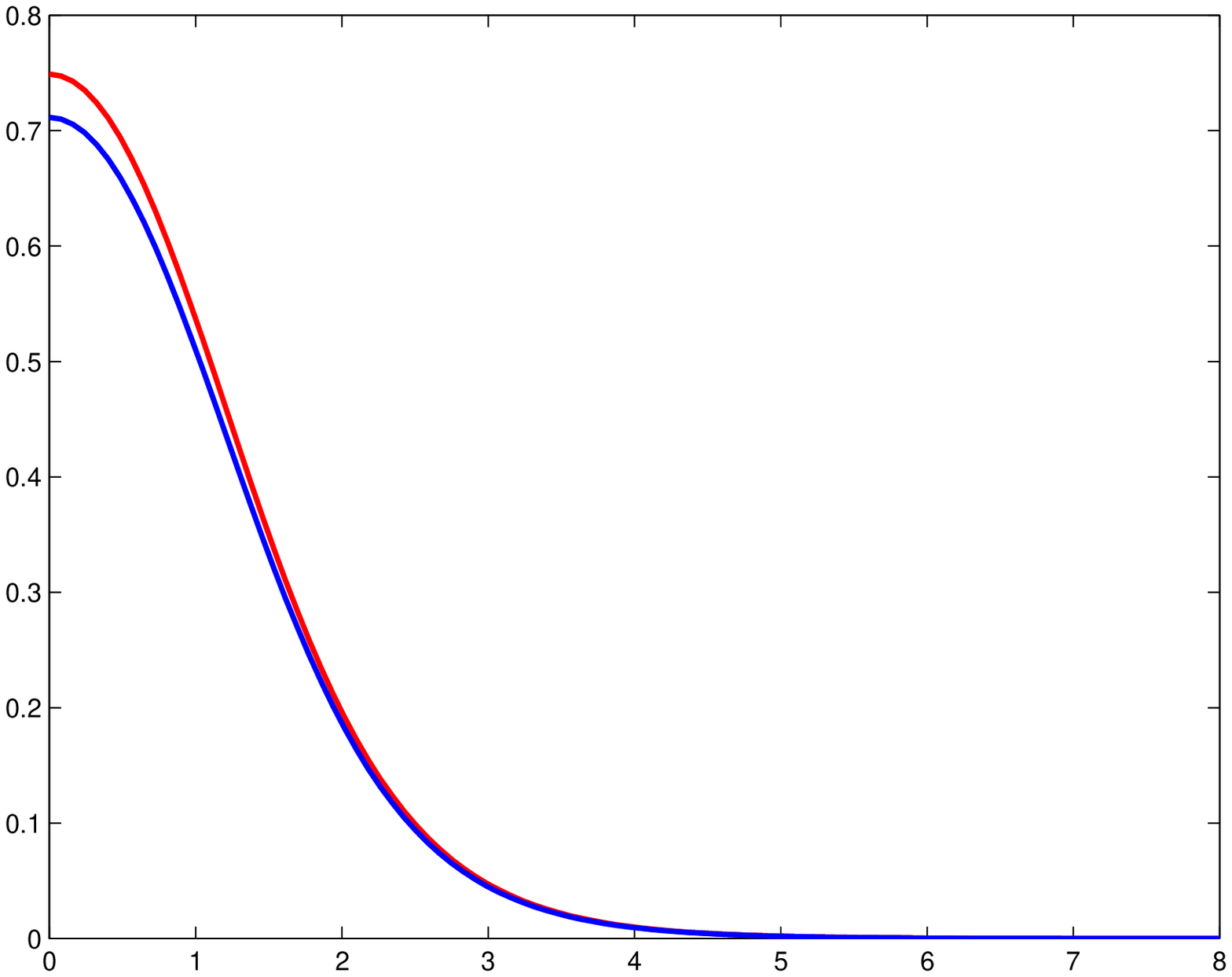} 
& 
\epsfxsize=60mm\epsfysize=60mm\epsfbox{\FigDir 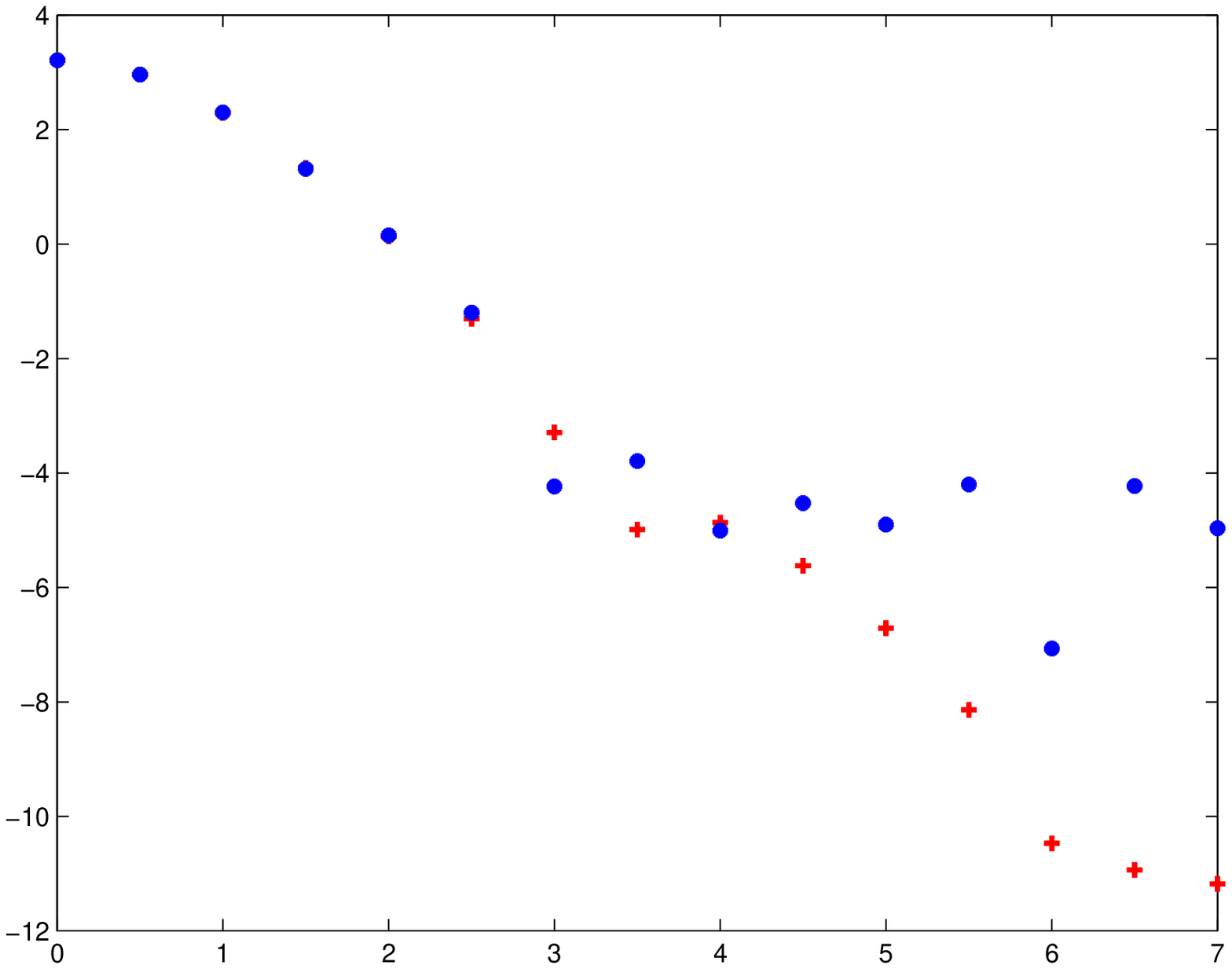} 
\end{tabular}
\end{tabular}
\\
\bigskip\centerline{\btabu{lll}
 Fig. 7: & a) $p(k,l|\yb)$;\\ 
			& b) $p(k|\yb)$ and $p(k|\yb,\lh)$; \\ 
			& c) $\rho(r)$ and $\wh{\rho}(r)$;\,   
			& d) $F(q_i)$ and $\wh{F}(q_i)$. 
\etabu}
\ecc

\section{Conclusions}
We discussed the different steps for a complete resolution of an inverse 
problem and focused on the choice of a basis function selection and the 
order of the model. An algorithm based on Bayesian estimation is proposed 
and tested on simulated data.  

%\newpage

\edoc